\documentclass[aps,superscriptaddress,twocolumn,twoside,floatfix,pra,nofootinbib,a4paper]{revtex4}
\usepackage{times}
\usepackage{epsfig}
\usepackage{amsfonts}
\usepackage{amsmath}
\usepackage{amssymb}
\usepackage{color}
\usepackage{multirow}

\usepackage[normalem]{ulem}
\usepackage{latexsym}
\usepackage{amsfonts}
\usepackage{mathrsfs}
\usepackage{natbib}
\usepackage{verbatim}
\usepackage{gensymb}
\usepackage[T1]{fontenc}

\usepackage[colorlinks=true,linkcolor=blue,citecolor=magenta,urlcolor=blue]{hyperref}

\DeclareMathOperator{\Tr}{tr}

\newcommand{\ket}[1]{|#1\rangle}
\newcommand{\bra}[1]{\langle#1|}

\newcommand{\norm}[1]{\left\lVert#1\right\rVert}

\begin{document}


\title{Self-testing non-projective quantum measurements in prepare-and-measure experiments}


\author{Armin Tavakoli}
\affiliation{D\'epartement de Physique Appliqu\'ee, Universit\'e de Gen\`eve, CH-1211 Gen\`eve, Switzerland}

\author{Massimiliano Smania}
\affiliation{Department of Physics, Stockholm University, S-10691 Stockholm, Sweden}

\author{Tam\'as V\'ertesi}
\affiliation{Institute for Nuclear Research, Hungarian Academy of Sciences, P.O. Box 51, 4001 Debrecen, Hungary}

\author{Nicolas Brunner}
\affiliation{D\'epartement de Physique Appliqu\'ee, Universit\'e de Gen\`eve, CH-1211 Gen\`eve, Switzerland}

\author{Mohamed Bourennane}
\affiliation{Department of Physics, Stockholm University, S-10691 Stockholm, Sweden}

\begin{abstract}	
Self-testing represents the strongest form of certification of a quantum system. Here we investigate theoretically and experimentally the question of self-testing non-projective quantum measurements. That is, how can one certify, from observed data only, that an uncharacterised measurement device implements a desired non-projective positive-operator-valued-measure (POVM). We consider a prepare-and-measure scenario with a bound on the Hilbert space dimension, which we argue is natural for this problem since any measurement can be made projective by artificially increasing the Hilbert space dimension. We develop methods for (i) robustly self-testing extremal qubit POVMs (which feature either three or four outcomes), and (ii) certify that an uncharacterised qubit measurement is non-projective, or even a genuine four-outcome POVM. Our methods are robust to noise and thus applicable in practice, as we demonstrate in a photonic experiment. Specifically, we show that our experimental data implies that the implemented measurements are very close to certain ideal three and four outcome qubit POVMs, and hence non-projective. In the latter case, the data certifies a genuine four-outcome qubit POVM. Our results open interesting perspective for strong ``black-box'' certification of quantum devices.
\end{abstract}


\maketitle


\section{Introduction}
Measurements in quantum theory were initially represented by complete sets of orthogonal projectors on a Hilbert space. Such measurements are standard in a multitude of applications. Nevertheless, in a modern understanding of quantum theory, measurements are described by positive-operator valued measures (POVMs), i.e., a set of positive semi-definite operators summing to identity. POVMs are the most general notion of a quantum measurement; all projective measurements indeed are POVMs, but not all POVMs need be projective.

Non-projective measurements are widely useful in both conceptual and applied aspects of quantum theory, as well as in quantum information processing. In several practically motivated tasks, they present concrete advantages over projective measurements. Non-projective measurements enhance estimation and tomography of quantum states \cite{DB98, Renes}, as well as entanglement detection \cite{Shang}. Furthermore, they allow for unambiguous state discrimination of non-orthogonal states \cite{USD1, USD2, USD3}, which would be impossible with projective measurements. They have also found applications in quantum cryptography \cite{Crypto1, Crypto2} and randomness generation \cite{Brask}. In addition, non-projective measurements can be used to maximally violate particular Bell inequalities \cite{VB10} (assuming a bound on the Hilbert space dimension), a fact that has been applied to improve randomness extraction beyond what is achievable with projective measurements \cite{APV16, GM17}.

In view of their diverse and growing applicability, it is important to develop tools for certifying and characterising non-projective measurements under minimal assumptions. The strongest possible form of certification involves a ``black-box'' scenario, where the quantum devices are a priori uncharacterised. Astonishingly, it is possible in certain cases to completely characterise both the quantum state and the measurements based only on observed data, which is referred to as ``self-testing'' \cite{MY}. A well-known example is that the maximal violation of the Clauser-Horne-Shimony-Holt (CHSH) Bell inequality \cite{CHSH69} implies (self-tests) a maximally entangled two-qubit state, and pairs of anti-commuting local projective measurements \cite{SW87, PR92, T93}. Self-testing can also be made robust to noise \cite{ST2, Jed,RUV13}.

However, for the purpose of characterising non-projective measurements in the black-box scenario, methods based on Bell inequalities encounter a challenge. Due to Neumark's theorem, every non-projective measurement can be recast as a projective measurement in a larger Hilbert space. That is, any non-projective measurement on a given system is equivalent to projective measurement applied to the joint state of the system and an ancilla of a suitable dimension, see e.g. \cite{Oszmaniec}. Since in the Bell scenario one usually considers no restriction on Hilbert space dimension, it is non-trivial to characterise a non-projective measurement based on a Bell inequality. While this is possible in theory (in the absence of noise) \cite{APV16}, it appears challenging in the more realistic scenario where the experiment features imperfections. A possible way to circumvent this problem is to consider a Bell scenario with quantum systems of bounded Hilbert space dimension. In particular, Refs \cite{GGG16,GM17} recently reported the experimental certification of a non-projective measurement in a Bell experiment assuming qubits. However, these experiments do not represent self-tests, as they certify the non-projective character of a measurement, but not how it relates to a specific target POVM.

Here we investigate the problem of self-testing non-projective measurements. We follow a different approach, by considering a prepare-and-measure scenario instead of a Bell scenario. We argue that this approach is well suited to the problem, and offers a natural framework for certifying and characterising non-projective measurements. Indeed, as the problem almost naturally involves an upper bound on the Hilbert space dimension of the quantum systems (as discussed above), the prepare-and-measure scenario is sufficient. In this case, as opposed to Bell experiments, there is no need to involve distant observers and entangled states. From a practical point of view this makes a very significant difference, as prepare-and-measure experiments are notably simpler to implement \cite{Hendrych, Ahrens, SE16, Lunghi, TH15, MTC18, AB14}. Moreover, prepare-and-measure scenarios are easier to analyse theoretically, which allows us to develop self-testing methods that are versatile and highly robust to noise.

In the first part of the paper, we present methods for characterising non-projective measurements. Firstly, we present a method for self-testing a targeted non-projective measurement in noiseless scenarios. Secondly, since noiseless statistics never occur in practice, we present methods for inferring a lower bound on the closeness of the uncharacterised measurement and a given target POVM, based on the observed noisy statistics; specifically, we lower-bound the worst-case fidelity between the real measurement and the ideal target one. Thirdly, we introduce a method for determining whether the observed statistics could have arisen from some (unknown) projective measurements. If not, the measurement is certified as non-projective. These methods have two-fold relevance. On the one hand, they enable foundational insights to physical inference of non-projective measurements in the black-box scenario. On the other hand, they provide tools for assessing and certifying the quality of an experimental setup. We demonstrate the practicality of these self-testing methods in two experiments. In the first, we target a symmetric informationally complete (SIC) qubit POVM and demonstrate an estimated $98\%$ worst-case fidelity in the black box scenario. Additionally, our data certifies a genuine four-outcome qubit POVM. In the second experiment, we target a symmetric three-outcome qubit POVM and certify a worst-case fidelity of at least $96\%$. Finally, we discuss some open questions.

\section{The self-testing problem, the scenario and overview of results}\label{overview}

Self-testing is the task of characterising a quantum system based only on observed data, i.e., black-box tomography. In other words, it is about gaining knowledge of the physical properties of initially unknown states and/or measurements present in an experiment by studying the correlations observed in the laboratory.

In this work, we focus on prepare-and-measure scenarios. They differentiate themselves from Bell scenarios in two important ways. Firstly, prepare-and-measure scenarios involve communicating observers and thus no space-like separation. Secondly, they do not involve entanglement, whereas Bell scenarios do. Prepare-and-measure scenarios can generally be modelled by two separated parties, Alice and Bob, who receive random inputs $x$ and $y$ respectively. Alice prepares and sends a quantum state $\rho_x$ to Bob who performs a measurement $y$ with outcome $b$, represented by a POVM $\{M_y^b\}_b$ with
\begin{align} M_y^b\geq0     \quad \text{and} \quad  \sum_b M_y^b=\openone  \quad \forall y.
\end{align}
 This generates a probability distribution 
 \begin{align}\label{prob}p(b|x,y)=\Tr\left(\rho_xM_y^b\right).
 \end{align} 
 
 In order to make the problem non-trivial, an assumption on Alice's preparations is required; otherwise Alice could simply send $x$ to Bob and any probability distribution $p(b|x,y)$ would be achievable. The assumption we consider in this work is that Alice's preparations, i.e. the set of states $\rho_x$, can be represented in Hilbert space of given dimension $d$. By choosing $d < |x|$, we prevent Alice from communicating all information about her input $x$ to Bob. Importantly, there exist distributions obtained from quantum systems of a dimension $d$ that cannot be simulated classically, see e.g. \cite{Gallego, BNV13}. That is, no strategy in which Alice communicates a classical $d$-valued message to Bob can possibly reproduce the observed data. Such distributions that cannot be classically simulated are candidates for self-testing considerations. 
 
 The problem of self-testing consists in characterising the set of states $\{\rho_x \}$ and/or the set of measurements $\{M_y^b\}$ based only on the distribution $p(b\lvert x,y)$. This characterisation can usually be done only up to a  unitary transformation and possibly a relabelling. In a recent work \cite{TKV18}, methods were presented for self-testing sets of pure quantum states, as well as sets of projective measurements in the qubit case. These were subsequently extended to higher dimensional systems in Ref.~\cite{Farkas}.

 Formally, a self-test can be made via a \textit{witness}, which is a linear function of the probability distribution $p(b\lvert x,y)$: 
 \begin{align}
  \mathcal{A}[p(b\lvert x,y)] = \sum_{x,y,b} \alpha_{xyb} p(b\lvert x,y) 
  \end{align} 
  where $\alpha_{xyb}$ are real coefficients. 
  Moreover, given a witness, one can determine its maximal witness value $\mathcal{A}^Q$ achievable under quantum distributions \eqref{prob} in a bounded Hilbert space. The witness can then be used for self-testing a set of quantum states and/or measurements, whenever there is a unique combination of states and/or measurements that achieves $\mathcal{A}^Q$. Then, it is clear that when the observed distribution $p(b\lvert x,y)$ leads to $\mathcal{A}^Q$, a specific set of states and/or measurements is identified (up to a simple class of transformations). A necessary condition for a witness to be useful for self-testing is that, for a given dimension $d$, quantum systems outperform classical ones; if not, several strategies would generally be compatible with the data; see Refs \cite{Gallego,Hendrych,Ahrens,PB11} for examples of such witnesses. In Section~\ref{selftest1}, we present a method for constructing witnesses whose maximal value can self-test a targeted non-projective qubit measurement $\mathcal{M}^{\text{target}}$.

\begin{figure}[b!]
	\centering
	\includegraphics[width=\columnwidth]{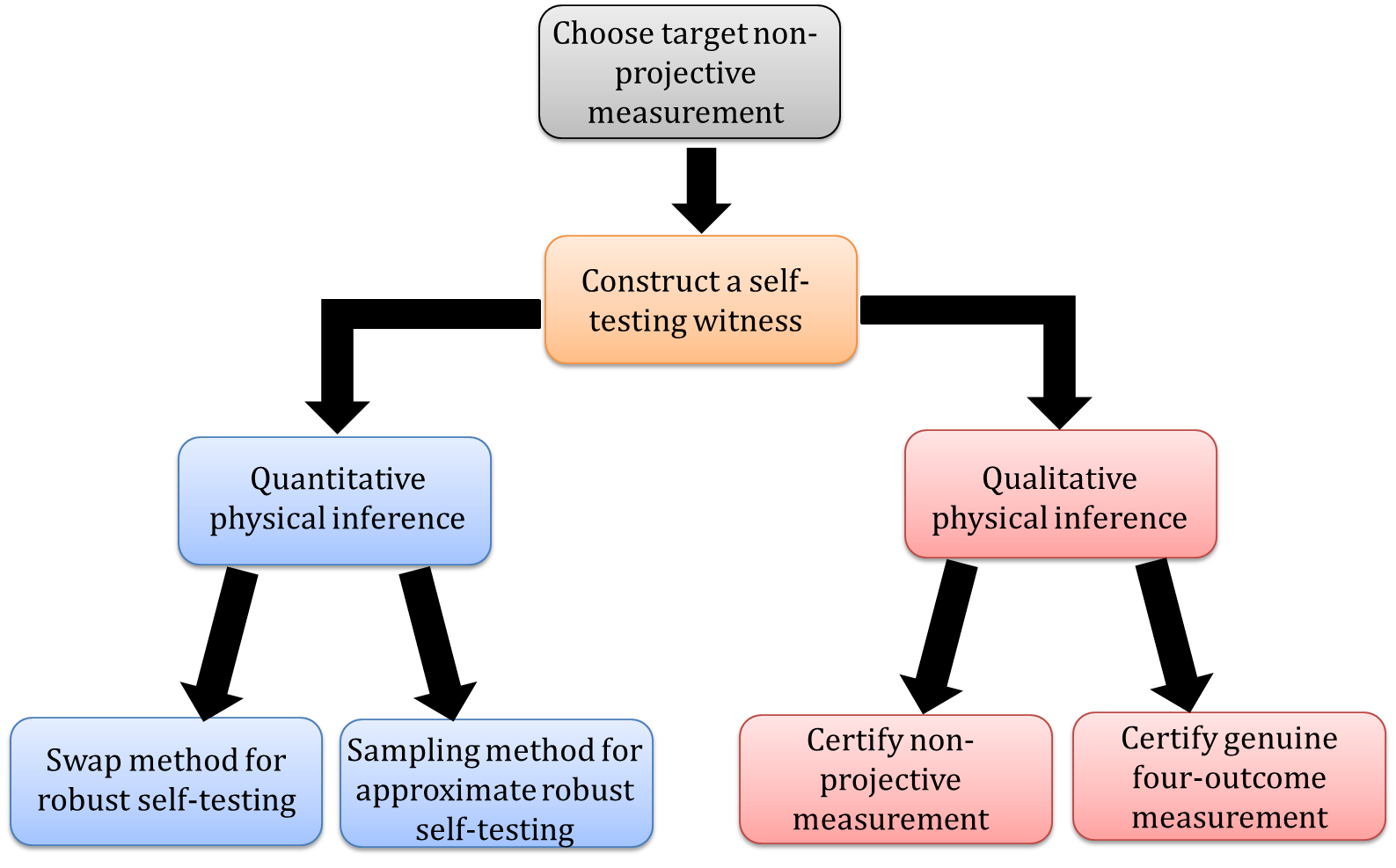}
	\caption{Graphical overview of the self-testing methods and steps presented in section~\ref{methods}.}\label{MethodsTreeFig}
\end{figure}

Next we turn to robust self-tests, i.e., self-tests that can be applied even when the statistics is not ideal causing the witness value to be less than  $\mathcal{A}^Q$. Indeed, this is fundamental in order to make our methods applicable in practice, as any realistic experiment is prone to noise. The influence of noise makes it impossible to perfectly pinpoint the states and measurements. This motivates the following question. Given an observation of a witness value $\mathcal{A}<\mathcal{A}^Q$, how close are the states and measurements to the ideal ones, i.e. those that would have been perfectly self-tested if we had observed $\mathcal{A}=\mathcal{A}^Q$. In Section.~\ref{selftest3}, we develop methods for robustly self-testing non-projective qubit measurements by lower-bounding the fidelity between the implemented measurement and the ideal one. A tight robust self-testing would give the fidelity between the measurement that is most distant from the ideal one, and that could have generated a witness value $\mathcal{A}<\mathcal{A}^Q$.

Whereas robust self-testing represents a quantitative physical inference, it is also relevant to consider a more qualitative inference. Based on the witnesses we develop for self-testing, we show how to certify that the uncharacterised measurement is non-projective. In section~\ref{selftest2} we determine the largest value of our witness that is compatible with qubit projective measurements. When observing a larger value, the non-projective character of the measurement is certified. In a similar spirit, we determine a bound on our witness above which a genuine four-outcome (non-projective) qubit measurement is certified.

An overview of all the self-testing methods developed in this work are illustrated in Fig.~\ref{MethodsTreeFig}. The methods will be applied in section~\ref{casestudies} to self-test particularly relevant non-projective qubit measurements. For these examples, we will demonstrate the usefulness of our methods by implementing them in a photonic experiment. Specifically, our experimental data implies that the implemented measurements are very close to certain ideal three- and four- outcome qubit POVMs, and hence are non-projective. In the latter case, the data certifies a genuine four-outcome qubit POVM.

\section{Certification and characterisation of non-projective qubit measurements}\label{methods}
This section presents methods for certification and characterisation of non-projective measurements in prepare-and-measure scenarios both with noiseless and noisy statistics. The focus will be on qubit systems. Therefore, we begin by summarising the properties of qubit POVMs.

A positive operator-valued measure (POVM) with $O$ outcomes is a set of operators $\{E_i\}_{i=1}^O$ with the property that $E_i\geq 0$ and that $\sum_i E_i=\openone$. In the case of qubits, $E_i$ can be represented on the Bloch sphere as 
\begin{equation}\label{povm}
E_i=\lambda_i\left(\openone+\vec{n}_i\cdot \vec{\sigma}\right),
\end{equation}
where $\vec{n}_i$ (with $\lvert \vec{n}_i\rvert\leq 1$) is the Bloch vector, $\lambda_i\geq 0$, and $\vec{\sigma}=\left(\sigma_x,\sigma_y,\sigma_z\right)$ are the Pauli matrices. Positivity and normalisation imply that
\begin{align}\label{POVM}
& \sum_{i=1}^O \lambda_i=1 & \text{and} && \sum_{i=1}^O \lambda_i\vec{n}_i=0.
\end{align}
The set of POVMs is convex, and a POVM is called extremal if it cannot be decomposed as a convex mixture of other POVMs. For qubits, extremal POVMs have either $O=2,3,4$ outcomes \cite{AP05}. In the case $O=2$, extremal POVMs are simply projective, whereas for $O=3$ and $O=4$ they are non-projective; an extremal three-outcome qubit POVM has three unit Bloch vectors in a plane, and an extremal four-outcome qubit POVM has four unit Bloch vectors of which no choice of three are in the same plane \cite{AP05}. An extremal qubit POVM is therefore characterised by its Bloch vectors. As the statistics of non-extremal POVMs can always be simulated by stochastically implementing extremal POVMs, it is clear that only extremal POVMs can be self-tested.

\subsection{Self-testing non-projective measurements: noiseless case}\label{selftest1}

\begin{figure}
	\centering
	\includegraphics[width=0.9\columnwidth]{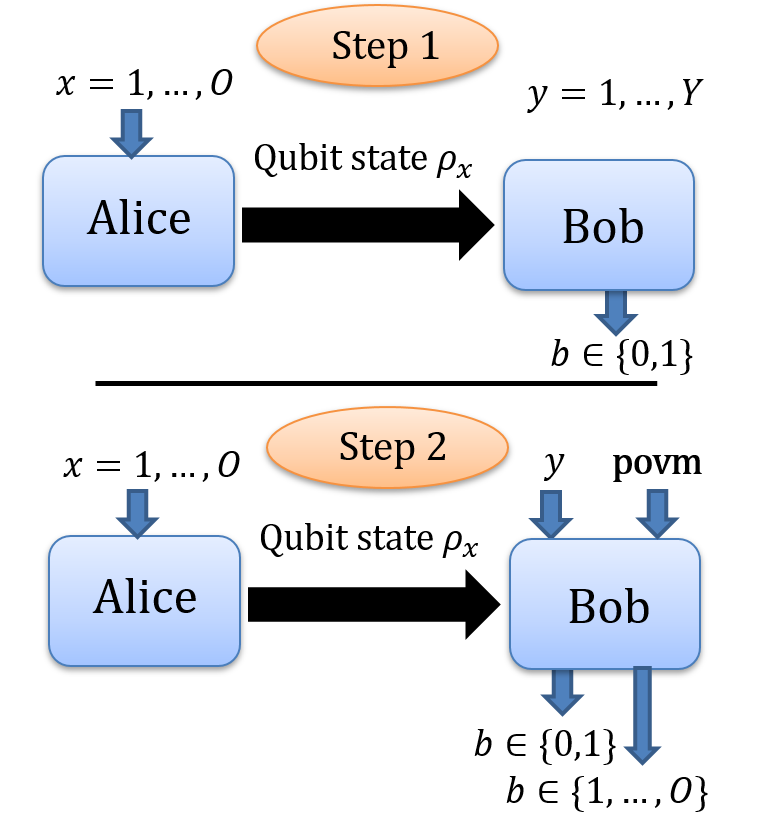}
	\caption{Method for self-testing a targeted non-projective qubit measurement by exploiting simpler self-tests of preparations. \textbf{Step 1}: tailor scenario and witness such that a maximal $\mathcal{A}'$ self-tests Alice's preparations to have Bloch vectors that are anti-aligned with those of the target measurement. \textbf{Step 2}: Add an extra setting to Bob and modify the witness to self-test the target non-projective measurement.}\label{methodfig}
\end{figure}

Consider a target extremal non-projective qubit POVM $\mathcal{M}^{\text{target}}$, with $O=3$ or $O=4$ outcomes, for which we associate the outcome $b$ to the unit Bloch vector $\vec{v}_b$. Our goal is now to construct a witness $\mathcal{A}$ such that its maximal value self-tests $\mathcal{M}^{\text{target}}$. The method consists in two steps summarised in Fig~\ref{methodfig}.

\textbf{Step 1.} First we construct a simpler witness $\mathcal{A}'$ featuring $O$ preparations, i.e. Alice has $O$ inputs. Bob receives an input $y=1,\ldots, Y$ and provides a binary outcome. The goal of this simpler witness is to self-test a particular relation among the prepared states $\ket{\psi_x}$. Specifically, we would like to certify that their unit Bloch vectors $\vec{u}_x$ point in opposite direction (on the Bloch sphere) to those of the target POVM $\mathcal{M}^{\text{target}}$, i.e. $\vec{u}_x = - \vec{v}_x$ for $x=1,...,O$. Let us define
\begin{equation}
\label{Aprime}
\mathcal{A}'=\sum_{x,y,b}c_{xyb} P(b|x,y),
\end{equation}
with real coefficients $c_{xyb}$ chosen such that the maximal value  $\mathcal{A}'^Q$ of the witness for qubits self-tests the desired set of prepared states $\{\ket{\psi_x}\}$ (up to a global unitary and relabellings). In general, we believe that it is always possible to find such a self-test by considering enough inputs for Bob, corresponding to well-chosen projective measurements, and suitable coefficients $c_{xyb}$; see Ref.\cite{TKV18} for examples. Furthermore, note that one could also in principle have more than $O$ preparations for Alice, and then self-test that  $O$ of them have the desired relation to $\mathcal{M}^{\text{target}}$. In addition, we remark that the construction of an adequate witness $\mathcal{A}'$ is not unique in general.

\textbf{Step 2.} We construct our final witness $\mathcal{A}$ from $\mathcal{A}'$. Specifically, we supply Bob with one additional measurement setting called $\mathbf{povm}$. This setting corresponds to a measurement with $O$ outcomes. Since the intention is to self-test the measurement corresponding to this setting as $\mathcal{M}^{\text{target}}$, we associate the setting $\mathbf{povm}$ to  $O$ outcomes.  We define
\begin{equation}\label{step2}
\mathcal{A}=\mathcal{A}'-k\sum_{x=1}^{O} P(b=x |x,\mathbf{povm}),
\end{equation}
for some positive constant $k$. A maximal witness value $\mathcal{A}^Q = \mathcal{A}'^Q$ now implies that the setting $\mathbf{povm}$ corresponds to $\mathcal{M}^{\text{target}}$ (up to a unitary and relabellings). This is because
a maximal witness value implies that (i) the set of prepared states $\{\ket{\psi_x}\}$ have Bloch vectors anti-aligned with those of $\mathcal{M}^{\text{target}}$, and (ii) $P(b=x |x,\mathbf{povm})=0$ for all $x$, hence the Bloch vectors of the setting $\mathbf{povm}$ are of unit length and aligned with those of $\mathcal{M}^{\text{target}}$. Moreover, as a qubit POVM is characterised by its Bloch vectors, we see that $\mathcal{M}^{\text{target}}$ is the only POVM that can attain the maximal witness value $ \mathcal{A}^Q$. Therefore we obtain a self-test of the target POVM $\mathcal{M}^{\text{target}}$.

In section~\ref{casestudies}, we will apply this method to self-test symmetric qubit POVMs with three and four outcomes.

\subsection{Robust self-testing of non-projective measurements}\label{selftest3}
No experiment can achieve the noiseless conditions needed to obtain \textit{exactly} a maximal value of $\mathcal{A}$. Therefore, it is paramount to discuss the case when a non-maximal value of $\mathcal{A}$ is observed. We will show that in this case, one can nevertheless make a statement about how close the uncharacterised measurement $\mathcal{E}$ performed in the lab (corresponding to the setting $\mathbf{povm}$) is to the target POVM $\mathcal{M}^{\text{target}}$.

In order to address this question, we must first define a measure of closeness between two measurements. A natural and frequently used distance measure in quantum information is the fidelity, $F$, between two operators. Inspired by previous works \cite{Jed,TKV18,MO, JD1}, we consider a measure of closeness amounting to the best possible weighted average fidelity between the extremal qubit target POVM elements $\mathcal{M}^{\text{target}}=\{M_i\}$ and the actual POVM elements $\mathcal{E}=\{E_i\}$. That is, we allow for a quantum extraction channel $\Lambda$ to be applied to the actual POVM. This extraction channel must be unital (i.e., identity preserving) in order to map a POVM to another POVM. Clearly, we look for the best possible extraction channel. We thus define the quantity
\begin{equation}
\label{fidave}
F\left(\mathcal{E},\mathcal{M}^{\text{target}}\right)=\max_{\Lambda} \frac{1}{2}\sum_{i=1}^{O}\frac{\Tr\left(\Lambda[E_i]M_i\right)}{\Tr\left(M_i\right)}.
\end{equation}
Since the target measurement is extremal, the POVM elements are proportional to rank-one projectors; $M_i\propto P_i$. Due to \eqref{povm} we can write $\Lambda[E_i]=\lambda_i\left(\openone+\vec{n}_i\cdot \vec{\sigma}\right)$ subject to the  constraints \eqref{POVM}. By evaluating \eqref{fidave} we find that $F=1/2+1/2\sum_i \lambda_i \Tr\left(P_i \vec{n}_i\cdot \vec{\sigma}_i\right)\leq 1$. To saturate the inequality, each Bloch vector $\vec{n}_i$ must be of unit length, i.e. $|\vec{n}_i|=1$, and aligned with the Bloch vector of $P_i$. Hence, $M_i$ and $\Lambda[E_i]$ are both proportional to the same rank-one projector. Since a POVM with Bloch vectors of unit length is fully characterised, i.e. all coefficients $\lambda_i$ are fixed by the conditions \eqref{POVM}, this implies that $M_i=\Lambda[E_i]$. Thus, a maximal fidelity of $F=1$ is uniquely achieved when the actual POVM is equal to the target measurement \footnote{We note that this would not necessarily be the case when the POVM $M_i$ is not extremal. However, as mentioned above, for the purpose of self-testing it is enough to focus on the case where $M_i$ is extremal.}.

In general, a non-maximal value of the witness $\mathcal{A}$ can arise from many different possible choices of states and measurements. We denote by $S(\mathcal{A})$ the set of all $O$-outcome POVMs that are compatible with a given observed value $\mathcal{A}$. Our goal is now to find a lower-bound on the average fidelity $F$ that holds for \textit{every} measurement $\mathcal{E}' \in S(\mathcal{A})$. Therefore, the quantity of interest is the \textit{worst-case} average fidelity:
\begin{equation}\label{fidlow}
\mathcal{F}\left(\mathcal{A}\right)=\min_{\mathcal{E}'\in S(\mathcal{A})}F\left(\mathcal{E}',\mathcal{M}^{\text{target}}\right).
\end{equation}
Calculating this quantity, or even lower-bounding it, is typically a non-trivial problem even in the simplest case. We proceed with presenting two methods for this task. 

We remark that the definition \eqref{fidave}, given for qubits, could potentially be extended to higher-dimensional systems (replacing the factor $1/2$ by $1/d$). This could work for POVMs where all elements are proportional to rank-one projectors. However, the latter are only a strict subset of general extremal POVMs. Finding a more general figure of merit is thus an interesting open question.

\textit{Robust self-testing with the swap-method.---} A lower-bound on the worst-case average fidelity can be obtained via semidefinite programming \cite{SDP}. The method combines the so-called swap-method \cite{YV14, BN15}, introduced for self-testing in the Bell scenario, and the hierarchy of dimensionally bounded quantum correlations \cite{NV15}. Such adaptations of the swap-method to prepare-and-measure scenarios were introduced in \cite{TKV18} to self-test pure state and projective measurements. In Appendix~\ref{appSwap} we outline the details of how the swap-method is adapted to robustly self-test non-projective measurements. This method benefits from being applicable in a variety of scenarios and for returning rigorous lower-bounds on $\mathcal{F}$. Nevertheless, it suffers from two drawbacks. Firstly, the method only overcomes the fact that self-tests are valid up to a global unitary, but not that they may be valid up to relabellings. Thus, it is only useful for target measurements that are self-tested up to a unitary. Secondly, while rarely producing tight bounds on $\mathcal{F}$, the computational requirements scale rapidly with the number of inputs, the number of outputs, and the chosen level of the hierarchy. In Section~\ref{casestudies}, we will show that the method can be efficiently applied for robustly self-testing a three-outcome qubit POVM.

\textit{Numerically approximating robust self-testing.---} In order to also address cases in which self-tests are valid up to both a unitary transformation and relabellings, we can estimate $\mathcal{F}$ based on random sampling. The approximation method benefits from being straightforward and broadly useful, while it suffers from the fact that it merely estimates the value of $\mathcal{F}$ instead of providing a strict lower-bound. The key feature is that the minimisation appearing in Eq.~\eqref{fidlow} is replaced by a minimisation taken over data obtained from many random samples of the setting $\mathbf{povm}$. We detail this method in Appendix~\ref{appApprox}, and apply it to an example in Section.~\ref{casestudies}.

\subsection{Certification methods for non-projective measurements}\label{selftest2}
Whereas robust self-testing considers quantitative aspects of physical inference from noisy data, it is important to also consider the qualitative inference. An important qualitative statement is to prove that the uncharacterised measurement is non-projective, or more generally, that it cannot be simulated by projective measurements. Indeed, it is known that when POVMs are sufficiently noisy, they become perfectly simulable via projective measurements \cite{Oszmaniec,Hirsch,Leo}. The witnesses we construct can address this question. We will see that whenever the observed value of the witness $\mathcal{A}$ is sufficiently large, one can certify that the setting $\mathbf{povm}$ necessarily corresponds to some non-projective measurement, and could not have been simulated via projective measurements. Specifically we derive an upper-bound on $\mathcal{A}$ for projective measurements (or convex combination of them). The violation of such a bound thus certifies a non-projective measurement, or more precisely a genuine three (or four) outcome POVM. At the end of this subsection, we also show how to certify a genuine four outcome POVM.

A projective qubit measurement has binary outcomes, and can therefore be represented by an observable $M\equiv M_0-M_1$ where $M_i$ is the measurement operator corresponding to outcome $i=0,1$. Let us consider the case where the $O$-outcome measurement $\mathbf{povm}$ is projective. One may assign two outcomes to rank-one projectors and the rest to trivial zero operators. Note that it is enough here to consider these cases, as the witness $\mathcal{A}$ is linear in terms of the measurement operators. Projectors can thus be assigned in three ($O=3$) or six ($O=4$) different ways, of which the optimal instance must be chosen. Let the outcomes in the optimal instance be  $o_{0|\mathbf{povm}}$ and $o_{1|\mathbf{povm}}$, and associate the observable $M_{\mathbf{povm}}\equiv M_{Y+1}=M_{\mathbf{povm}}^{o_{0|\mathbf{povm}}}-M_{\mathbf{povm}}^{o_{1|\mathbf{povm}}}$. The witness \eqref{step2} can be written as
\begin{equation}\label{pvmform}
\mathcal{A}=C(k)+\sum_{x} \Tr\left[\rho_{x}\mathcal{L}_{x}^{(k)}(\{M_y\})\right],
\end{equation}
where $C(k)$ is a constant, and $\mathcal{L}_x^{(k)}\left(\{M_y\}\right)$ is a linear combination of the observables $\{M_1,\ldots,M_{Y+1}\}$. Note that that $\mathcal{L}_x^{(k)}(\{M_y\})$ does not depend on the index $y$ but on the collection of observables.  Using the Cauchy-Schwarz inequality for operators we obtain,
\begin{equation}\label{pvmCS}
\mathcal{A}\leq C(k)+\sum_{x} \sqrt{\Tr\left[\rho_{x}\mathcal{L}_{x}^{(k)}(\{M_y\})^2\right]}.
\end{equation}
Due to projectivity, we have $M_y=\vec{n}_y\cdot\vec{\sigma}$, where $\vec{n}_y$ is of unit length. Using $\{M_k,M_l\}=2\vec{n}_k\cdot \vec{n}_l\openone$, one finds $\mathcal{L}_{x}^{(k)}(\{M_y\})^2= t_{x}^{(k)}\left(\{\vec{n}_y\}\right) \openone$, for some function $t$ which is a weighted sum of scalar products of the Bloch vectors of the observables. Consequently, in order to bound $\mathcal{A}$ under all projective measurements, we have
\begin{equation}\label{pvmbound}
\mathcal{A}\stackrel{\text{Proj}}{\leq}  C(k)+\max_{\{\vec{n}_y\}}\sum_{x}\sqrt{t_{x}^{(k)}\left(\{\vec{n}_y\}\right)}\equiv  \mathcal{B}(k).
\end{equation}
Thus, $\mathcal{B}(k)$ bounds the value of $\mathcal{A}$ for projective measurements. The evaluation of this bound only depends on Bob's Bloch vectors and is further simplified by their parameterisation in terms of two angles\footnote{The effort needed to evaluate the bound depends on the chosen prepare-and-measure scenario. Typically, considering scenarios with some symmetry properties is beneficial.}.

Moreover, when targeting a four-outcome qubit POVM, we consider also a finer form of qualitative characterisation by considering whether $\mathcal{A}$ can be simulated by the setting $\mathbf{povm}$ being some three-outcome POVM. If not, the measurement is certified as a genuine four-outcome measurement. This amounts to bounding the value of $\mathcal{A}$ achievable under any two- or three-outcome qubit POVM and then observing a violation of that bound. For this purpose, one may employ the hierarchy of dimensionally bounded quantum correlations \cite{NV15} which can be used to upper bound $\mathcal{A}$ under three-outcome POVMs \footnote{The hierarchy is built on projective measurements. This obstacle can be overcome by embedding Alice's preparations in a larger Hilbert space with the dimension chosen such that three-outcome POVMs can be re-cast as projective measurement following Neumark's theorem.}. To obtain tight bounds, one may need a reasonably high hierarchy level which can be efficiently implemented using the methods of Ref.~\cite{TRR18}.

Next, in Section.~\ref{casestudies}, we will apply the outlined methods to specific non-projective measurements and experimentally demonstrate the certification of both non-projective and genuine four-outcome measurements.

\section{Relevant examples and their experimental realisation}\label{casestudies}
In the above, we have discussed methods for self-testing a target non-projective measurement. Here we put these methods in practice in a photonic experiment. We implement three- and four-outcome symmetric qubit POVMs, with Bloch vectors forming a star (trine-POVM) and a tetrahedron (SIC-POVM) respectively. In the first case we certify a non-projective measurement, and apply our methods for robust self-testing, demonstrating worst-case average fidelity of at least $96\%$ compared to an ideal trine-POVM. In the second case, we certify a genuine four-outcome qubit POVM, and demonstrate worst-case average fidelity of approximately $98\%$ with respect to an ideal SIC-POVM. Below we will first present the setup common to both experiments and then consider each example separately by first applying the methods of section~\ref{methods} to obtain adequate witnesses, and then present the corresponding experimental realisation. 

\begin{figure*}
	\includegraphics[width=0.75\textwidth]{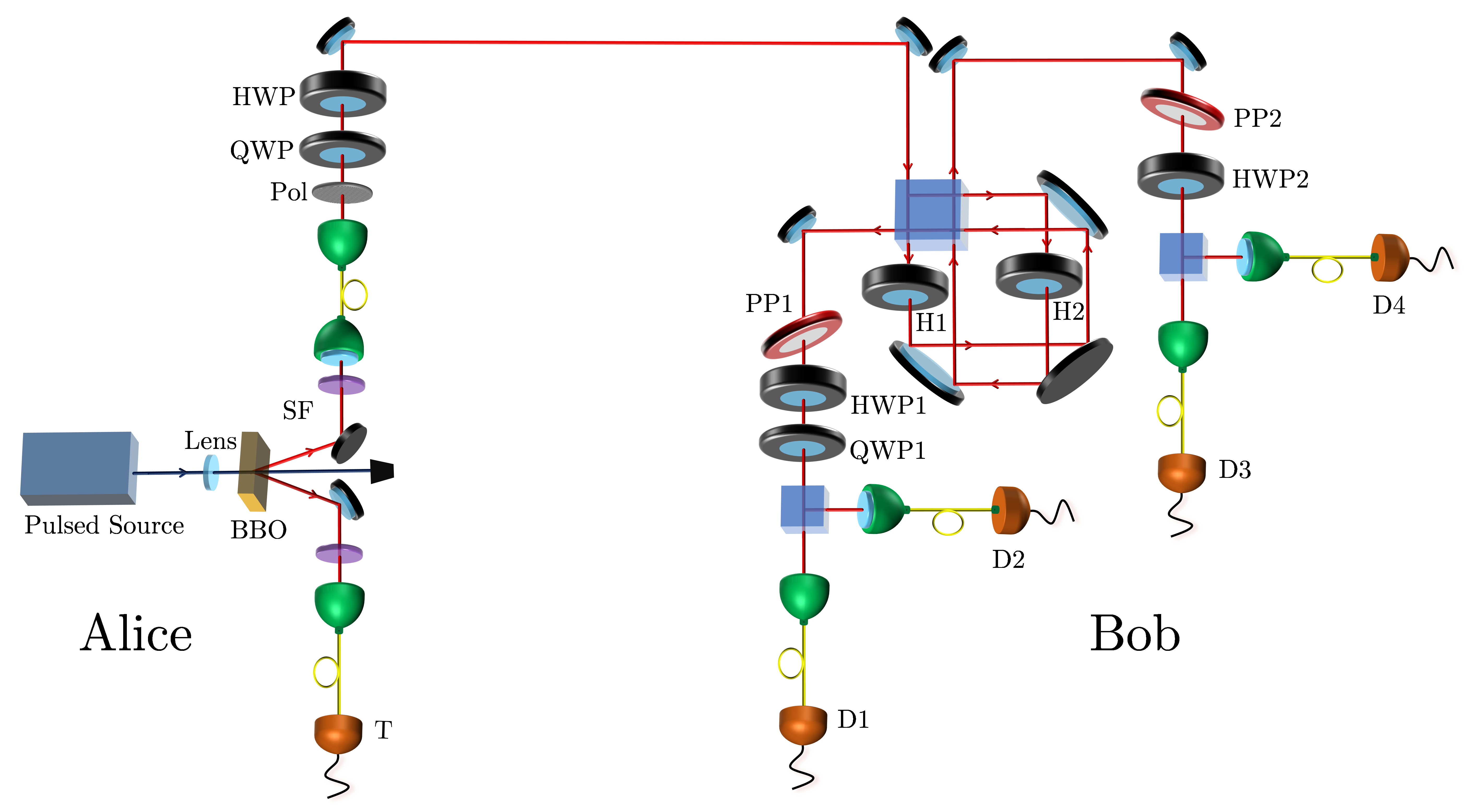}
	\caption{\label{fig:setup}%
		Experimental setup. More details, including labelling, can be found in the main text.}
\end{figure*}

\textit{Experimental setup --} 
In the experiment, the qubit states are encoded in the polarisation degree of freedom of a single photon, with the convention of $\ket{H} \equiv \ket{0}$ and $\ket{V} \equiv \ket{1}$. The setup is depicted in Fig.~\ref{fig:setup}. 
Alice's station includes a heralded single photon source where femto-second laser pulses at 390~nm are converted into pairs of photons at 780~nm, through type-I spontaneous
parametric down-conversion in two orthogonally oriented beta-barium borate crystals (BBO). Photon pairs go through 3~nm spectral filters (SF), and are then coupled into two single-mode fibres (SMF) for spatial mode filtering. The idler photon is sent to the trigger APD detector (T), and heralds the presence of a signal photon. The latter is then emitted again into free-space, and undergoes Alice's state preparation, consisting of a fixed linear polariser (POL), a $\lambda /4$ (QWP) and a $\lambda /2$ (HWP) wave-plate.

Upon preparing the required qubit state, Alice forwards the signal photon to Bob's measurement station, where it goes through a double-path Sagnac interferometer, each path of which contains an HWP. The interferometer mixes the polarisation degree of freedom with path, effectively enabling Bob to perform either projective or non-projective measurements in the original polarisation Hilbert space where the qubit was prepared, thanks to the two polarisation analysers at the outputs. Each of these consists in a phase plate (PP), a HWP and (in output 1) a QWP, a polarising beam-splitter and two single-photon detectors.
Outputs from all detectors (T, D1-D4) are sent to a coincidence unit connected to a computer.

All measurements were performed with heralded photon rates of approximately $1\times10^{4}$ counts per second, while each setting was measured for 500 seconds.
The quality of state preparation and measurement can be estimated by preparing states $\ket{H}$, $\ket{+}=(\ket{H}+\ket{V}) / \sqrt{2}$ and $\ket{R}=(\ket{H}+i\ket{V}) / \sqrt{2}$, and measuring them in the Pauli bases $\sigma_z$, $\sigma_x$ and $\sigma_y$ respectively.
The three visibilities obtained in our setup with this characterisation measurement were:
\begin{align}\nonumber
\mathcal{V}_{\sigma_z} &= (99.91 \pm 0.02) \% \\
\mathcal{V}_{\sigma_x} &= (99.31 \pm 0.01) \% \\\nonumber
\mathcal{V}_{\sigma_y} &= (99.23 \pm 0.02) \% .
\end{align}

While the almost optimal $\mathcal{V}_{\sigma_z}$ is a direct consequence of the high extinction ratios of the PBSs used, the lower visibilities in the interference bases are mainly due to the double-path Sagnac interferometer, which showed a visibility of around $99.4 \%$, therefore effectively bounding from above the results we can achieve in the experiments.


\subsection{Example I: the qubit SIC-POVM}\label{exampleSIC}
We begin by illustrating the self-testing methods for a frequently used non-projective measurement, namely the qubit SIC-POVM, which we denote $\mathcal{M}_{\text{SIC}}$. This measurement has four outcomes and its four unit Bloch vectors $\{\vec{v}_b\}_b$ form a regular tetrahedron on the Bloch sphere, with weights $\lambda_b=1/4$. Such a regular tetrahedron construction can be achieved via two different labellings of the four outcomes that are not equivalent under unitary transformations. Up to a unitary transformation, each such SIC-POVM can be written with Bloch vectors
\begin{equation}\label{POVM_vectors}
\begin{aligned} 
\vec{v}_1&=[1, 1, 1]/\sqrt{3}  \qquad\qquad \vec{v}_2=[1, -1, -1]/\sqrt{3}\\
\vec{v}_3&=[-1, 1, -1]/\sqrt{3}  \qquad\ \, \vec{v}_4=[-1, -1, 1]/\sqrt{3},
\end{aligned}
\end{equation}
and the set of Bloch vectors $\{-\vec{v}_l\}_l$ respectively. 

\begin{figure}
	\includegraphics[width=0.9\columnwidth]{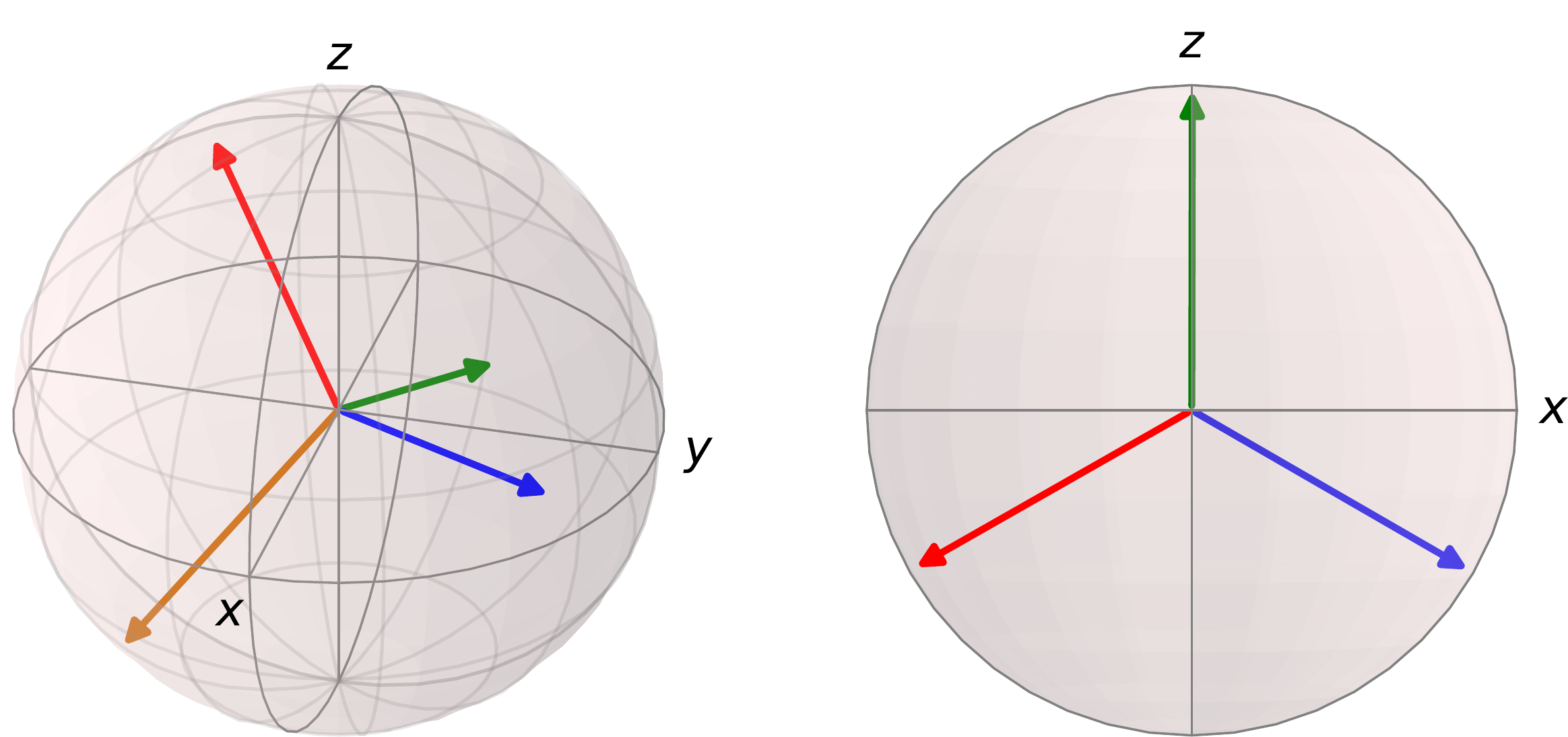}
	\caption{States prepared by Alice for the SIC-POVM (left) and trine-POVM (right) experiments. The vectors are obtained from qubit state tomography.}\label{bloch_both}
\end{figure}

\textit{Noiseless self-test.---} 
We find a prepare-and-measure scenario for self-testing $\mathcal{M}_{\text{SIC}}$. Following step 1 in section~\ref{selftest1}, we introduce a prepare-and-measure scenario in which Alice has four preparations, $x\in\{1,2,3,4\}$, and Bob has three binary-outcome measurements, $y\in\{1,2,3\}$. The witness is chosen as
\begin{equation}
\mathcal{A}'_{\text{SIC}}=\frac{1}{12}\sum_{x,y}P(b=S_{x,y}|x,y),
\end{equation}
where $S_{1,y}=\left[0, 0, 0\right]$, $S_{2,y}=\left[0, 1, 1\right]$, $S_{3,y}=\left[1, 0, 1\right]$, and $S_{4,y}=\left[1, 1, 0\right]$. The maximal value, $\mathcal{A}'_{\text{SIC}}=1/2\left(1+1/\sqrt{3}\right)$, can be achieved by Alice preparing her four states forming a regular tetrahedron, e.g., with the Bloch vectors in Eq. \eqref{POVM_vectors}, and Bob performing the measurements $\sigma_x,\ \sigma_y$ and $\sigma_z$. The four vectors experimentally prepared by Alice, as obtained by state tomography, are reported in Fig. \ref{bloch_both} (left).
In Appendix~\ref{PerfectExamples}, we prove the maximal witness value and show that it self-tests that Alice's preparations indeed must form a regular tetrahedron on the Bloch sphere. By step 2 in section~\ref{selftest1}, we supply Bob with an additional four-outcome measurement $\mathbf{povm}$, and consider the modified witness
\begin{equation}\label{SICexperiment}
\mathcal{A}_{\text{SIC}}=\frac{1}{12}\sum_{x,y}P(b=S_{x,y}|x,y)-k\sum_{x=1}^4P(b=x|x,\mathbf{povm}).
\end{equation}
Thus, we conclude that $\mathcal{A}_{\text{SIC}}=1/2(1+1/\sqrt{3})$ self-tests $\mathcal{M}_{\text{SIC}}$.

We note that there also exists other prepare-and-measure scenarios fulfilling the requirements of step 1. For example, one may achieve the desired self-test using the so-called $3\rightarrow 1$ random access code whose self-testing properties were considered in Ref.~\cite{TKV18}. However, this prepare-and-measure scenario requires more preparations than the one presented here.

\textit{Robust self-test --} 
Next, we consider the worst-case  fidelity (given in Eq.~\eqref{fidlow}) of the measurement corresponding to the setting $\mathbf{povm}$ with $\mathcal{M}_{\text{SIC}}$. Since the self-test of $\mathcal{M}_{\text{SIC}}$ is valid up to a relabelling as well as a collective unitary,  we cannot use the swap-method to lower bound $\mathcal{F}$. Instead, we employ the numerical approximation method (see Appendix~\ref{appApprox} for details). Figure~\ref{SICfig} displays roughly $3\times 10^5$ optimal pairs $(\mathcal{A}_{\text{SIC}},F)$ each evaluated from a randomly sampled measurement for the setting $\mathbf{povm}$. The evaluation was done for $k=1/5$ (which, as will soon be shown, turns out to be the most noise-resilient choice of $k$). We see that the minimal sampled fidelity as a function of $\mathcal{A}_{\text{SIC}}$  describes a curve which constitutes the approximation of $\mathcal{F}$.

\textit{Certifying non-projective  and genuine four-outcome POVMs.---} 
Finally, we derive a tight bound valid for all qubit projective measurements on the value of $\mathcal{A}_{\text{SIC}}$. Due to the symmetries of $\mathcal{A}_{\text{SIC}}$, we can without loss of generality let the non-trivial (non-zero measurement operator) outcomes of the measurement $\mathbf{povm}$ be the outcomes $b=1,2$. Hence, we define the observable $M_{\mathbf{povm}}\equiv M_4=M_{\mathbf{povm}}^1-M_{\mathbf{povm}}^2$. Then, we follow the steps outlined in section~\ref{selftest2}. First, we re-write $\mathcal{A}_{\text{SIC}}$ in the form \eqref{pvmform}. We find $C(k)=(1-2k)/2$ and
\begin{align}\nonumber
& \mathcal{L}_{x=0,1}^{(k)}(\{M_y\})=\frac{1}{24}\left[1,(-1)^{x},(-1)^x,(-1)^{x+1}12k\right]\cdot \vec{M} \\
& \mathcal{L}_{x=2,3}^{(k)}(\{M_y\})=\frac{1}{24}\left[-1,(-1)^{x},(-1)^{x+1},0\right]\cdot \vec{M},
\end{align}
where $\vec{M}=\left[M_1,M_2,M_3,M_4\right]$, with $M_y=\vec{n}_y\cdot \vec{\sigma}$. After applying the Cauchy-Schwarz inequality, we obtain a cumbersome expression of the form of Eq.~\eqref{pvmCS}. In order to evaluate its maximal value (following Eq.~\eqref{pvmbound}), we use the following concavity inequality: $\sqrt{r}+\sqrt{s}\leq \sqrt{2(r+s)}$ for $r,s\geq 0$, with equality if and only if $r=s$. Apply this inequality twice to the expression \eqref{pvmbound}, first to the two terms associated to $x=0,1$, and then to the two terms associated to $x=2,3$. After a simple optimisation over $\vec{n}_3$ and denoting $x=\vec{n}_1\cdot \vec{n}_2$, one arrives at
\begin{multline}\nonumber
\mathcal{A}_{\text{SIC}}\leq \frac{1-2k}{2}+\frac{\sqrt{2}}{24}\sqrt{6-4x}  \\
+\frac{\sqrt{2}}{24}\sqrt{2r_k+4x+48k\sqrt{2}\sqrt{1+x}}\equiv f_k(x),
\end{multline}
where $r_k=3+144k^2$. This bound is valid for a particular value of $x$. In order to hold for all projective measurements, we simply maximise $f_k(x)$ over $x$. This requires only an optimisation in a single real variable $x\in[-1,1]$ which is straightforward. The optimal choice is denoted $x^*$. Setting $\mathcal{B}(k)=f_k(x^*)$, we have $\mathcal{A}_{\text{SIC}} \leq \mathcal{B}(k)$ for all projective measurements. Although the expressions involved are cumbersome, the analysis is simple and straightforward. We have considered the tightness of the projective bound for $k\in\{1/100,2/100,\ldots, 1\}$ by numerically optimising $\mathcal{A}_{\text{SIC}}$ under unit-trace measurements (which includes all rank-one projective measurements). In all cases, we saturate the bound $\mathcal{B}(k)$ up to machine precision with a projective measurement.

\begin{figure}
	\includegraphics[width=0.9\columnwidth]{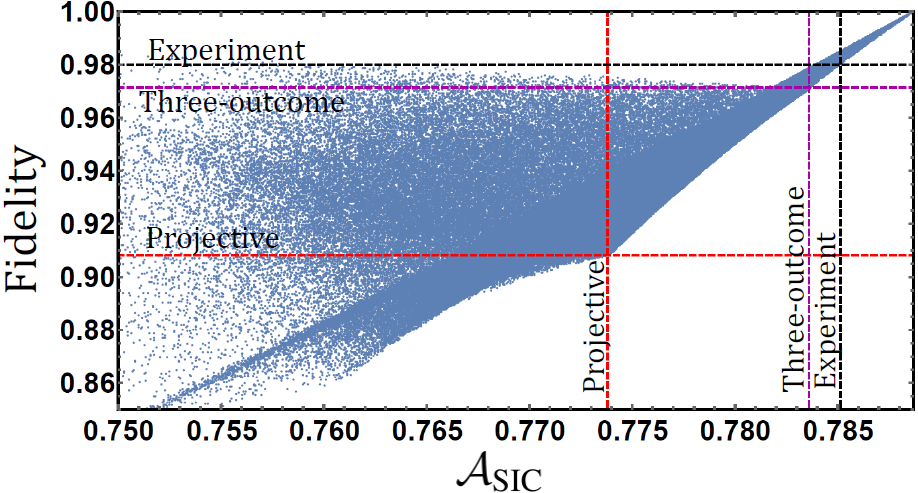}
	\caption{Numerical approximation of the worst-case  fidelity of the unknown measurement (setting $\mathbf{povm}$) with the qubit SIC-POVM by roughly $3\times 10^5$ random three- and four-outcome POVM samples for which the optimal values of $(\mathcal{A},F)$ were calculated. The figure also displays the critical limits on $\mathcal{A}_{\text{SIC}}$ and $\mathcal{F}$ for projective and three-outcome POVMs respectively, as well as the experimentally measured values.}\label{SICfig}
\end{figure}

Furthermore, we have also considered bounding $\mathcal{A}_{\text{SIC}}$ under three-outcome qubit POVMs using the hierarchy of dimensionally bounded quantum correlations (as described in section~\ref{selftest2}). In our implementation of \cite{NV15}, we have embedded the qubit preparations into a three-dimensional Hilbert space, and optimised $\mathcal{A}_{\text{SIC}}$ under projective measurements of the only existing non-trivial rank-combination. The relaxation level involved some monomials from both the second and third level, and the size of the moment matrix was $126$. This was done for all $k\in\{1/100,2/100\ldots,1\}$, and each upper bound was saturated up to numerical precision using lower bounds numerically obtained via semidefinite programs.

\begin{figure}
	\centering
	\includegraphics[width=\columnwidth]{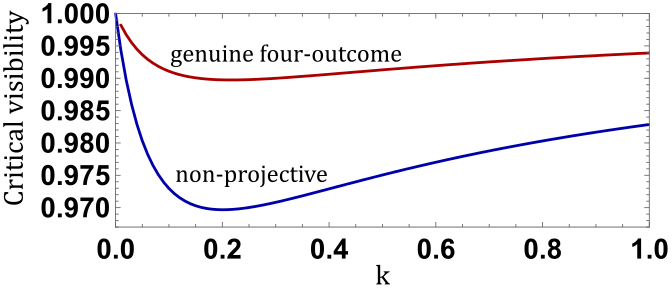}
	\caption{Critical visibilities for certifying a non-projective measurement and a genuine four-outcome measurement respectively, in the prepare-and-measure scenario \eqref{SICexperiment} targeting the qubit SIC POVM.}\label{Figvisi}
\end{figure}

In order to study the robustness of both the non-projective and the genuine four-outcome certification, we have considered the critical visibility of the system needed when exposed to white noise. This is modelled by the preparations taking the form $\rho_x(v)=v\rho_x+(1-v)\openone/2$ where $v\in[0,1]$ is the visibility. Denoting by $\mathcal{A}^{\text{rand}}$ the witness value obtained from the optimal measurements performed on the maximally mixed state, the critical visibility for violating some given bound $\mathcal{B}$ is 
\begin{equation}\label{visi}
v_{\text{crit}}(k)=\frac{\mathcal{B}(k)-\mathcal{A}^{\text{rand}}+k}{\mathcal{A}^{Q}-\mathcal{A}^{\text{rand}}+k}.
\end{equation}
We have applied this to $\mathcal{A}_{\text{SIC}}$ with $\mathcal{B}(k)$ corresponding to the bounds on projective and three-outcome measurements respectively. The corresponding critical visibilities appear in Figure~\ref{Figvisi}. In both cases, we find that the largest amount of noise is tolerated for $k=1/5$, corresponding to $v_{\text{crit}}=0.970$ and  $v_{\text{crit}}=0.990$ respectively.

\textit{Experimental result --} 
Wave-plate settings (referring to Fig. \ref{fig:setup}) for Alice's prepared states in Eq. \eqref{POVM_vectors} and Bob's measurements $\sigma_x,\ \sigma_y,\ \sigma_z$, and the four-outcome SIC-POVM anti-aligned to the vectors in Eq. \eqref{POVM_vectors}, are reported in Tab. \ref{tab:tetra_settings}.


\begin{table}[t!]
	\caption{\label{tab:tetra_settings}%
		Wave plate settings for the experimental setup (as appearing in Fig. \ref{fig:setup} ) for the experiment based on the SIC-POVM. All angles are in degrees. The preparation Bloch vectors $\vec{n}_i$ of Alice point to the vertices of a regular tetrahedron.}
	\begin{ruledtabular}
		\begin{tabular}{ccc}
			\multicolumn{3}{c}{Alice}\\
			\colrule
			state Bloch vector & HWP & QWP\\
			$\vec{n}_1$ & 20.07 & 17.63 \\
			$\vec{n}_2$ & 24.93 & -17.63 \\
			$\vec{n}_3$ & -24.93 & 17.63 \\
			$\vec{n}_4$ & -20.07 & -17.63 \\
		\end{tabular}
		\begin{tabular}{cccccccc}
			\multicolumn{7}{c}{Bob}\\
			\colrule
			setting & H1 & H2 & HWP1 & QWP1 & HWP2 & PP1 & PP2\\
			$1$ & 0 & 0 & 22.5 & 0 & -- & 0 & --\\
			$2$ & 0 & 0 & 0 & 45 & -- & 0 & --\\
			$3$ & 0 & 0 & 0 & 0 & -- & 0 & -- \\
			\textbf{povm} & 13.68 & 31.32 & 0 & 45 & 22.5 & 45 & 135 \\
		\end{tabular}
	\end{ruledtabular}
\end{table}
Optimally choosing $k=1/5$, the measured value of the witness as compared to the relevant bounds is
\begin{align}\nonumber
&\mathcal{A}_{\text{SIC}}\stackrel{\text{projective}}{\leq} 0.7738\stackrel{\text{3-outcome}}{\leq}0.7836  \stackrel{\text{qubit}}{\leq} 0.7887.\\
& \mathcal{A}_{\text{SIC}}^{\text{Lab}}=  0.78514 \pm 5 \times 10^{-5}_{stat} \pm 1.0 \times 10^{-4}_{syst}.
\end{align}
The statistical error originates from Poissonian statistics and the systematic error originates from the precision of the wave plate settings. More details about the errors are discussed in Appendix~\ref{ExperimentDetails}. 

We observe a substantial violation of both the projective measurement and the three-outcome measurement bounds. 
Thus, we can certify that Bob's measurement $\mathbf{povm}$ is a genuine four-outcome qubit POVM. Furthermore, as illustrated by the results in Fig~\ref{SICfig}, we certify approximately a $98\%$ worst-case  fidelity with the qubit SIC-POVM. 


\subsection{Example II: the qubit trine-POVM}\label{exampleTriangle}
We consider a second example in which the target POVM is the so-called trine-POVM. This measurement has three outcomes and its Bloch vectors form an equilateral triangle on a disk of the Bloch sphere, with $\lambda_l=1/3$. The Bloch vectors are hence defined by  
\begin{align}\label{vecs}
& \vec{v}_1=\left[0,0,-1\right] && \vec{v}_2=\frac{1}{2}\left[-\sqrt{3},0,1\right] && \vec{v}_3=\frac{1}{2}\left[\sqrt{3},0,1\right].
\end{align}

\textit{Noiseless self-test --} 
We introduce a prepare-and-measure scenario in which Alice has three inputs $x\in\{1,2,3\}$, and Bob has two binary-outcome measurements labelled by $y\in\{1,2\}$, and consider the witness
\begin{equation}
\mathcal{A}'_{\text{tri}}=\sum_{x,y,b}T_{x,y}(-1)^{b}P(b\lvert x,y),
\end{equation}
where $T_{x,1}=\left[1,1,-1\right]$ and $T_{x,2}=\left[\sqrt{3},-\sqrt{3},0\right]$. In Appendix~\ref{PerfectExamples}, we show that its maximal value is $\mathcal{A}'_{\text{tri}}=5$ and that this value implies that Alice's three preparations form an equilateral triangle on the Bloch sphere. Then, we add an additional input $\mathbf{povm}$ for Bob and consider the witness
\begin{equation}\label{tripovm}
\mathcal{A}_{\text{tri}}=\sum_{x,y,b}T_{x,y}(-1)^{b}P(b\lvert x,y)-k\sum_{x=1}^3 P(b=x|x,\mathbf{povm}),
\end{equation}
for some $k>0$. Then, $\mathcal{A}_{\text{tri}}=5$ self-tests the setting $\mathbf{povm}$ as the trine-POVM up to a unitary.

\begin{figure}
	\centering
	\includegraphics[width=\columnwidth]{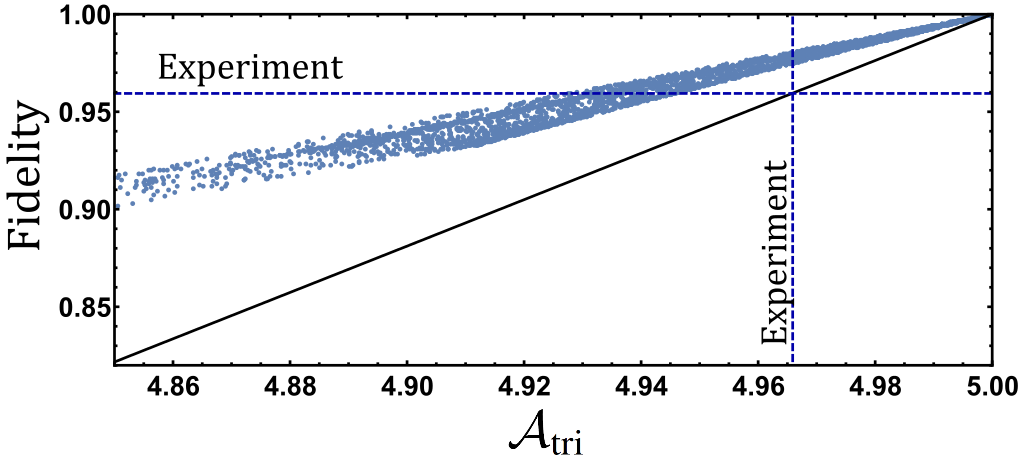}
	\caption{Lower bound on $\mathcal{F}(\mathcal{A}_{\text{tri}})$ for $k=1$ obtained from the swap method, together with roughly 3000 points $(\mathcal{A}_{\text{tri}},F)$ obtained via the numerical approximation method. This is displayed next to the experimentally achieved results.}\label{Figrobust}
\end{figure}

\textit{Robust self-test --} 
We now turn to considering its robust self-testing properties, i.e., lower-bounding the worst-case  fidelity of the unknown measurement (setting $\mathbf{povm}$) with the target measurement for a given value of $\mathcal{A}_{\text{tri}}$. Since the above self-test is achieved only up to unitary transformations, we may find rigorous lower bounds on the worst-case fidelity $\mathcal{F}$ using semidefinite programming. In accordance with section~\ref{selftest3}, we have performed the swap-operation on Bob's side and used the hierarchy of finite-dimensional correlations to lower bound $\mathcal{F}$ \footnote{The hierarchy level was an intermediate level containing some higher-order moments corresponding to an SDP matrix of size $105$.}. In addition, for sake of comparison, we have implemented the numerical approximation method for robust self-testing to estimate the accuracy of the bound obtained via the swap-method. The results are displayed in Figure~\ref{Figrobust}. Although the bound obtained from the swap-method is most likely not tight, it will prove sufficient for the practical purpose of experimentally certifying the targeted POVM with high accuracy.

Finally, we have also self-tested the trine-POVM in a different prepare-and-measure scenario (see Appendix~\ref{PerfectExamples}). In Appendix~\ref{ExamplesCertification}, we use this prepare-and-measure scenario to derive a tight bound on projective measurements by evaluating  the right-hand-side of \eqref{pvmbound}.

\textit{Experimental realisation --} 
The witness in Eq.~\eqref{tripovm} is maximised if Alice's three Bloch vectors point to the vertices of an equilateral triangle on a disk of the Bloch sphere. We take that disk to be the $xz$-plane  (see Fig. \ref{bloch_both} (right)), taking $\vec{t}_i= -\vec{v}_i$ (from \eqref{vecs}), and Bob performs one of three measurements $\sigma_z$, $\sigma_x$ and the three-outcome POVM with vectors anti-aligned to Alice's states. In contrast to the previous experiment,  the output 2 of Bob's measurement station only consists of one detector (D3), and no wave-plate or PBS (see Fig.~\ref{fig:setup}).
The wave-plate settings corresponding to the above states and measurements are reported in Tab. \ref{tab:tri_settings}.

\begin{table}[h]
	\caption{\label{tab:tri_settings}%
		Wave plate settings for the experimental setup (as appearing in Fig. \ref{fig:setup} ) for the experiment based on the trine-POVM. All angles are in degrees. The preparation Bloch vectors $\vec{t}_i$ of Alice point to the vertices of an equilateral triangle.}
	\begin{ruledtabular}
		\begin{tabular}{ccc}
			\multicolumn{3}{c}{Alice}\\
			\colrule
			state Bloch vector & HWP & QWP\\
			$\vec{t}_1$ & 0 & 0 \\
			$\vec{t}_2$ & 30 & 0 \\
			$\vec{t}_3$ & -30 & 0 \\
		\end{tabular}
		\begin{tabular}{cccccccc}
			\multicolumn{7}{c}{Bob}\\
			\colrule
			setting & H1 & H2 & HWP1 & QWP1 & HWP2 & PP1 & PP2\\
			$1$ & 0 & 0 & 0 & 0 & -- & 0 & -- \\
			$2$ & 0 & 0 & 22.5 & 0 & -- & 0 & -- \\
			\textbf{povm} & 0 & 27.37 & 22.5 & 0 & -- & 0 & -- \\
		\end{tabular}
	\end{ruledtabular}
\end{table}

Using these settings, we have obtained the experimentally measured value of $\mathcal{A}_{\text{tri}}$ as a function of $k$.  Since we aim to demonstrate a large worst-case  fidelity with the trine-POVM, we have computed the lower bound on $\mathcal{F}(\mathcal{A}_{\text{tri}})$ for many different values of $k$ and found that choosing $k=1$ leads to the optimal result. The corresponding experimentally measured witness is
\begin{align}
&  \mathcal{A}_{\text{tri}}(k=1)\stackrel{\text{projective}}{\leq} 4.89165  \stackrel{\text{qubit}}{\leq} 5\\
& \mathcal{A}_{\text{tri}}^{\text{Lab}}(k=1) = 4.96587 \pm 7 \times 10^{-4}_{stat} \pm 1.7 \times 10^{-3}_{syst}.
\end{align}
This data-point, and its relation to the worst-case  fidelity of the lab measurement with the targeted POVM is depicted in Fig.~\ref{Figrobust}. From $\mathcal{A}_{\text{tri}}^{\text{Lab}}$, we infer a closeness of at least  $96\%$. This can be compared to the largest possible fidelity between a projective measurement and the trine-POVM, which is straightforwardly found to be $(2+\sqrt{3})/4\approx 0.933$. However, as is indicated by the results of the sampling based numerical approximation method for robust self-testing (presented in Fig.~\ref{Figrobust}), a better bound of $\mathcal{F}$ may allow us to rigorously infer a worst-case  fidelity of at least $97.3\%$.

Furthermore, we have considered the possibility of the experimental data certifying a non-projective qubit measurement. However, to this end, we found that another choice of $k$ is optimal with respect to the witness value that is achievable under projective measurements. We found that the optimal choice is $k\approx 4.5$. The corresponding experimentally measured value becomes
 \begin{align}\nonumber
 &\mathcal{A}_{\text{tri}}(k=4.5)\stackrel{\text{projective}}{\leq} 4.71139  \stackrel{\text{qubit}}{\leq} 5\\
 & \mathcal{A}_{\text{tri}}^{\text{Lab}}(k=4.5)=  4.93613 \pm 5 \times 10^{-5}_{stat} \pm 1.0 \times 10^{-4}_{syst}.
 \end{align} 
We conclude that our experimental data certifies a non-projective qubit measurement.

\section{Conclusions}
We investigated the problem of self-testing non-projective measurements. We argued that a prepare-and-measure scenario with an upper bound on the Hilbert space dimension represents a natural framework for investigating this problem. We considered both the qualitative certification of a measurement being non-projective and/or genuine four-outcome, as well as a quantitative characterisation in terms of worst-case fidelity to a given target POVM. We demonstrate the practical relevance of these methods in two experiments in which we both certify a genuine four-outcome POVM, and infer a high worst-case fidelity with respect to target symmetric qubit POVMs.

It would be interesting to overcome the limitation of the swap-method and develop a rigorous robust self-testing method for general four-outcome qubit POVMs. Also extending these methods to high-dimensional POVMs would be relevant since there exist extremal non-projective measurements that feature the same number of outcomes as projective measurements (contrary to the qubit case).  Moreover, it would be interesting to investigate self-testing of non-projective measurements using different assumptions as in our work. One could consider for instance prepare-and-measure scenarios with a bound on the entropy \cite{Chaves}, the overlap between the prepared states \cite{Brask} or their mean energy \cite{Tom}. Finally, one may ask whether it would be possible to robustly self-test a non-projective measurement in the fully device-independent case, i.e. returning to the Bell scenario without any assumption on the dimension. 

\emph{Note added.---} During the completion of this manuscript, we became aware of an independent work \cite{Piotr} discussing the certification of qubit POVMs.

\section{Acknowledgements}
We thank J\k{e}drzej Kaniewski for insightful comments. This work was supported by the Swiss national science foundation (Starting grant DIAQ, NCCR-QSIT), the Swedish research council, and Knut and Alice Wallenberg Foundation. T.V. is supported by the National Research, Development and Innovation Office NKFIH (Grant Nos. K111734 and KH125096).

\newpage
\onecolumngrid
\appendix
\newpage

\section{Swap-method for robust self-testing of non-projective measurements}\label{appSwap}
We outline the swap-method for robust self-testing of non-projective qubit measurements. Bob is supplied with trusted ancillary qubits
\begin{equation}\label{barrhox}
\bar{\rho}_b=\ket{\bar{\psi}_b}_{\text{B'}}\bra{\bar{\psi}_b},
\end{equation}
where $\text{B'}$ denotes the ancillary system of Bob. The key idea is that Bob will swap $\bar{\rho}_b$ into his measurement box via an operation depending on his uncharacterised measurements, and then perform the de facto measurement (determined by the inputs $y$ and $\mathbf{povm}$) on the swapped system. 

Let us recall our figure of merit from Eq.~(\ref{fidave}), where $\mathcal{M}^{\text{target}}=\{M_b\}_b$ is the target measurement. Without loss of generality, we can let every POVM element $M_b$ be proportional to a rank-1 projector with a proportionality factor $2\lambda_b$ (see Eq.~(\ref{POVM}) for the definition of $\lambda_b$)\footnote{ Note that in Sec.~\ref{casestudies}, the presented case studies will focus on $\lambda_b=1/d$, where $d$ is the number of outcomes.}. The ancillary qubits are initialised such that  $M_b=2\lambda_b\bar{\rho}_b$ corresponding to the ideal case in which a noiseless self-test can be made. We may write the worst-case fidelity in Eq.~(\ref{fidlow}) as
\begin{equation}\label{FA1}
\mathcal{F}(\mathcal{A})=\min_{\{M_b\}\in S(\mathcal{A})}\max_{\Lambda}\frac{1}{2}
\sum_b\Tr\left[\Lambda[M_{b}]\bar{\rho}_b\right],
\end{equation}
Next, we obtain a lower bound on this quantity by using the swap method adapted to POVMs in the prepare-and-measure scenario (e.g. see Ref.~\cite{TKV18} for self-testing state preparations in these setups). To this end, $\bar{\rho}_b$ in Eq.~(\ref{barrhox}) is swapped into Bob's box. Hence, in Eq.~(\ref{FA1}) we replace $\bar{\rho}_b$ by
\begin{equation}
\label{rhoswap}
\rho_b^{\text{SWAP}}=\Tr_{\text{B'}}\left[\text{SWAP}(\rho_b\otimes\bar{\rho}_b)\text{SWAP}^{\dagger}\right],
\end{equation}
where $\text{SWAP}=UVU$ is the swap operator with
\begin{align}
\label{swap}
U&=\openone \otimes \ket{0}_{\text{B'}}\bra{0}+B_1\otimes\ket{1}_{\text{B'}}\bra{1},\nonumber\\
V&=\frac{1+B_0}{2}\otimes \openone_{\text{B'}}+\frac{1-B_0}{2}\otimes ({\sigma_x})_{\text{B'}},
\end{align}
where $B_0$ and $B_1$ are for Bob's qubit observables corresponding to projective two-outcome measurements. Note that in the ideal case, one has $B_0=\sigma_z$ and $B_1=\sigma_x$, and then $\text{SWAP}$ defines the two-qubit swap operator
\begin{equation}
\text{SWAP}=\sum_{i,j=0}^1\ket{i}\bra{j}\otimes\ket{j}_{\text{B'}}\bra{i}.
\end{equation}
However, if Bob's $B_0$ and $B_1$ measurements which give rise to the maximal witness value $\mathcal{A'}$ in Eq.~(\ref{Aprime}) are not equivalent up to unitary rotations to $B_0=\sigma_z$ and $B_1=\sigma_x$, we can still construct the swap operator. We can always rotate Bob's local bases such that $B_1=\sigma_x$ and $B_0=\sin\theta \sigma_z + \cos\theta \sigma_x$ for some $\theta$ angle. In order to derive the optimal operator used to construct the swap operator, we can naively set $B_z=(B_0-\cos\theta B_1)/\cos\theta$. However, this operator needs not be unitary in general when it is expressed in terms of the moments. Still there always exist a unitary $B_2$ such that $B_2^{\dagger}B_z\geq 0$. We then use $B_2$ in the construction of the swap instead of $B_0$, which ensures the unitarity of the swap operator. More on this so-called localising matrix approach can be found in Refs.~\cite{YV14, BN15}.

By replacing $\bar{\rho}_b$ with (\ref{rhoswap}) in Eq.~(\ref{FA1}), we obtain
\begin{align}\nonumber\label{FA2}
\mathcal{F}(\mathcal{A})=\min_{\{M_{b}\}\in S(\mathcal{A})}\max_{\Lambda}
\sum_b\frac{1}{2} \Tr\left[\Lambda[M_{b}]\otimes\openone \text{SWAP}(\rho_b\otimes\bar{\rho}_b)\text{SWAP}^{\dagger}\right].
\end{align}
We can further write $\text{SWAP}=(1/2)\sum_{i,j=0}^1{s_{ij}\otimes\ket{i}_{\text{B'}}{_\text{B'}}\bra{j}}$, where $s_{ij}$ are some linear combinations of polynomial expressions in $B_0$, $B_1$ and $\openone$, whose exact forms can be found in Appendix H of Ref.~\cite{TKV18}. Using the above decomposition for $\text{SWAP}$ one arrives at
\begin{equation}\label{FA22}
\mathcal{F}(\mathcal{A})=\min_{\{M_{b}\}\in S(\mathcal{A})}
\frac{1}{8}\sum_b\sum_{i,j,k=0}^1 \Tr\left(R_{i,j,k,b}\right)\bra{j}\bar{\rho}_b\ket{k},
\end{equation}
where $R_{i,j,k,b}=M_{b}s_{ij}\rho_b s_{ki}$.  Note that the channel $\Lambda$ maps measurements to measurements, therefore we safely ignored it in the above expression.

Lower bounds valid for any quantum realisation of the  \eqref{FA22} can be obtained via the hierarchy of Ref.~\cite{NV15}. This is a finite dimensional SDP method using a random sampling approach to construct a  feasible moment matrix $\Gamma$  with entries $\Gamma_{i,j}=\Tr{\left(Q_i^{\dagger}Q_j\right)}$, where $Q_i$ defines a sequence of operators such that $\Gamma$ contains all variables $R_{i,j,k,b}$ and also contains all variables building up $\mathcal{A}$. Then the solution of the following SDP provides a lower bound to $\mathcal{F}(\mathcal{A^*})$, where $\mathcal{A^*}$ stands for a fixed value of $\mathcal{A}$:
\begin{align}\label{sdp}
\mathcal{F}(\mathcal{A}^*)\geq \frac{1}{32}\min_{\Gamma}\sum_{b}\sum_{i,j,k=0}^{1} \Tr\left(R_{i,j,k,b}\right)\langle j\lvert \bar{\rho}_{b}\lvert k\rangle \\\nonumber
\text{such that } \hspace{3mm}\Gamma\geq 0 \hspace{2mm} \text{  and  } \hspace{2mm} \mathcal{A}\geq \mathcal{A}^*.
\end{align}

Note that during the random sampling, we generate quantum states $\ket{\psi_b}$ in $\mathbb{C}^2$ along with traceless qubit observables $B_y$ in $\mathbb{C}^2$. However, in order to simulate a random $d$-outcome measurement $\mathcal{M}$ in $\mathbb{C}^2$, one has to generate random projective measurements in a larger space, and then embed the preparations $\ket{\psi_b}$ and observables $B_y$ in this larger space. Consequently, a feasible moment matrix $\Gamma$ is constructed from these quantum states and measurements~\cite{NV15}.

\section{Approximation method for robust self-testing of non-projective measurements}\label{appApprox}
In this appendix, we detail a simple procedure for estimating $\mathcal{F}$ in \eqref{fidlow} for robust self-testing. The key idea is to replace the minimisation in \eqref{fidlow} by instead sampling a large number of elements from the relevant set of POVMs.

First, sample a random extremal $d$-outcome qubit POVM corresponding to the setting $\mathbf{povm}$. Numerically compute the maximal value of $\mathcal{A}_{\text{SIC}}$ over the $O$ preparations and the $Y$ binary-outcome measurements. For qubits, this can be straightforwardly achieved using Bloch sphere parameterisation and the fact that all variable measurements are optimally taken as projective. An alternative method is to use semidefinite programs in see-saw \cite{seesaw}. Then, optimise $F$ over the unital extraction channel $\Lambda$. A simple way of obtaining a lower bound on $F$ is to relax $\Lambda$ to a unitary, which benefits from a simple Bloch sphere parameterisation. Perform this optimisation separately with respect to each of the possible relabellings of the target measurement that are not equivalent under unitary transformations, and then choose the largest one. In this manner, one obtains an optimised pair $(\mathcal{A},F)$ for the given measurement sample for the setting  $\mathbf{povm}$. Repeating this procedure many times, and collect a large sample of pairs $(\mathcal{A},F)$. Then, one can estimate the minimisation over the set $S(\mathcal{A})$ appearing in Eq.~\eqref{fidlow} by instead, for each $\mathcal{A}$ taken in a suitably small interval, choosing the smallest $F$ among the obtained pairs $(\mathcal{A},F)$. In order to have a good approximation of $\mathcal{F}$, one requires a reasonably large number of samples as well as a reasonably unbiased manner of sampling the measurement $\mathbf{povm}$.

\section{Self-testing  non-projective measurements with noiseless statistics}\label{PerfectExamples}
In this Appendix, we exemplify the method for self-testing a targeted non-projective measurement in various cases. First, we focus on the four-outcome qubit SIC-POVM discussed in the main text. It is defined up to a  unitary and relabellings, by Bloch vectors $\vec{v}_{r_0r_1}=1/\sqrt{3}\left[(-1)^{r_0},(-1)^{r_1},(-1)^{r_0+r_1}\right]$, and weights $\lambda_{r_0r_1}=1/4$, for $r_0,r_1\in\{0,1\}$. Then, we consider the three-outcome trine-POVM also discussed in the main text. It is defined up to a unitary, by Bloch vectors $\vec{v}_1=\left[1,0,0\right]$, $\vec{v}_2=\left[-1,0,\sqrt{3}\right]/2$ and $\vec{v}_3=\left[-1,0,-\sqrt{3}\right]/2$, and weights $\lambda_r=1/3$, for $r\in\{1,2,3\}$. For these two cases, we first use the corresponding prepare-and-measure scenarios described in the main text and prove the desired self-tests. Then, we prove self-testing of the trine-POVM in a different type of prepare-and-measure scenario exhibiting practical symmetries.

\subsection{Self-testing the SIC-POVM}
In the main text, an example of a prepare-and-measure scenario is given in which a maximal witness value self-tests a qubit SIC-POVM. Here, we provide all details for the discussion in the main text.

Provide Alice with a random input $x\in \{1,2,3,4\}$ which is associated to a qubit, and Bob with a random input $y\in\{1,2,3\}$, and consider the following witness
\begin{equation}
\mathcal{A}'_{\text{SIC}}=\frac{1}{12}\sum_{x,y} P(b=S_{x,y}\lvert x,y),
\end{equation}
where $S_{1,y}=\left[0, 0, 0\right]$, $S_{2,y}=\left[0, 1, 1\right]$, $S_{3,y}=\left[1, 0, 1\right]$, and $S_{4,y}=\left[1, 1, 0\right]$. In a quantum model, we can use that $M_y^0+M_y^1=\openone$ to write
\begin{equation}\label{tetra1}
\mathcal{A}_{\text{SIC}}'=\frac{1}{2}+\frac{1}{12}\sum_{y}\Tr\left[M_y^0W_y\right],
\end{equation}
where $W_y=\sum_x (-1)^{S_{x,y}}\rho_x$. We must show that a maximal value of $\mathcal{A}'_{\text{SIC}}$ implies that Alice's preparation Bloch vectors form a regular tetrahedron on the Bloch sphere. To this end, the Cauchy-Schwarz inequality gives
\begin{equation}\label{tetra}
\sum_{y}\Tr\left[M_y^0W_y\right]\leq \sum_{y}\sqrt{\Tr\left[M_y^0W_y^2\right]\Tr\left[M_y^0\right]}.
\end{equation}
However, all extremal two-outcome measurements are projective, and must necessarily be rank-one for qubits since they otherwise lead to trivial statistics. Therefore, $\Tr\left[M_y^0\right]=1$.

To further simplify the problem, we use the following concavity inequality: for $q_k\geq 0$,
\begin{equation}\label{conc}
\left(\sum_{k=1}^N\sqrt{q_k}\right)^2\leq N\sum_{k=1}^N q_k,
\end{equation}
with equality if and only if all $q_k$ are equal. Taking $q_y=\Tr\left[M_y^0W_y^2\right]$ we obtain
\begin{equation}\label{tetra2}
\sum_{y}\sqrt{\Tr\left[M_y^0W_y^2\right]} \leq \sqrt{3\sum_{y}\Tr\left[M_y^0W_y^2\right]}.
\end{equation}
Using Bloch sphere representation, we write $\rho_x=1/2\left(\openone+\vec{m}_x\cdot \vec{\sigma}\right)$, for $|\vec{m}_x|\le 1$, and optimally align $M_y^0$ with $W_y^2$ to obtain
\begin{align}\label{m1}
\sum_{y}\Tr\left[M_y^0W_y^2\right]&=\frac{1}{4}\left[\left(\vec{m}_1+\vec{m}_2-\vec{m}_3-\vec{m}_4\right)^2+\left(\vec{m}_1-\vec{m}_2+\vec{m}_3-\vec{m}_4\right)^2+\left(\vec{m}_1-\vec{m}_2-\vec{m}_3+\vec{m}_4\right)^2\right]\\
&=\frac{1}{4}\left[3\sum_x|\vec{m}_x|^2-2\sum_{i<j} \vec{m}_i\cdot\vec{m}_j\right].
\end{align}
Since the second term on the right-hand-side can be positive, the maximal value necessitates pure states ($|\vec{m}_x|=1$). We must find the set of unit Bloch vectors that minimises $\sum_{i<j} \vec{m}_i\cdot\vec{m}_j$:
\begin{multline}
\min \left(\sum_{i<j} \vec{m}_i\cdot\vec{m}_j\right) = \min  \left(-\norm{\vec{m}_2+\vec{m}_3+\vec{m}_4}+\vec{m}_2\cdot\left(\vec{m}_3+\vec{m}_4\right)+\vec{m}_3\cdot\vec{m}_4\right)\\
=\min \{-\sqrt{3+2\left(\vec{m}_2\cdot \vec{m}_3+\vec{m}_2\cdot \vec{m}_4+\vec{m}_3\cdot \vec{m}_4\right)}+\vec{m}_2\cdot \vec{m}_3+\vec{m}_2\cdot \vec{m}_4+\vec{m}_3\cdot \vec{m}_4\} =-2,
\end{multline}
where in the first step we have optimally anti-aligned $\vec{m}_1$ and $\vec{m}_2+\vec{m}_3+\vec{m}_4$, and in the last step we have solved $\frac{d}{dx}\left(-\sqrt{3+2x}+x\right)=0$ with $x=\vec{m}_2\cdot \vec{m}_3+\vec{m}_2\cdot \vec{m}_4+\vec{m}_3\cdot \vec{m}_4$. Note that this implies that the maximal $\mathcal{A}'_{\text{SIC}}$ is $\mathcal{A}_{\text{SIC}}'= 1/2(1+1/\sqrt{3})$. The condition for minimality is therefore
\begin{equation}\label{mid1}
\vec{m}_2\cdot\vec{m}_3+\vec{m}_2\cdot\vec{m}_4+\vec{m}_3\cdot\vec{m}_4=-1.
\end{equation}
We can without loss of generality use the parameterisation $\vec{m}_2=[0,0,1]$, $\vec{m}_3=[\sin\epsilon,0,\cos\epsilon]$ and $\vec{m}_4=\left[\sin\theta\cos\phi,\sin\theta\sin\phi,\cos\theta\right]$. Solving the equation in $\phi$, the solution of interest is
\begin{equation}
\phi=\arccos\left(-\cot\left(\frac{\epsilon}{2}\right)\cot\left(\frac{\theta}{2}\right)\right).
\end{equation}
However, in order for our upper bound in \eqref{tetra2} to be tight, we need $\Tr\left[M_0^0W_0^2\right]=\Tr\left[M_1^0W_1^2\right]=\Tr\left[M_2^0W_2^2\right]$. Hence, we require that   $N_1=\norm{\vec{m}_1+\vec{m}_2-\vec{m}_3-\vec{m}_4}^2$, $N_2=\norm{\vec{m}_1-\vec{m}_2+\vec{m}_3-\vec{m}_4}^2$ and $N_3=\norm{\vec{m}_1-\vec{m}_2-\vec{m}_3+\vec{m}_4}^2$ are  equal. With the given parameterisation, these reduce to $N_1=-8\left(\cos \epsilon+\cos\theta\right)$, $N_2=8\left(1+\cos\theta\right)$ and $N_3=8\left(1+\cos\epsilon\right)$. To solve  $N_2=N_3$, we need $\cos\epsilon=\cos\theta$, and then the equation $N_1=N_2$ reduces to  $1+3\cos\theta=0$. The solution is $\theta=\pm\arccos(-1/3)$. Thus, we have obtained the desired solution. Plugging the solution back into $\vec{m}_i$, it is then easily seen that $\vec{m}_j\cdot\vec{m}_k=-1/3$ for $j\neq k$ which defines a regular tetrahedron, which is the desired relation between Alice's preparations.

By step 2, we supply Bob with one additional setting, $\mathbf{povm}$, and construct the new witness
\begin{equation}\label{tetragame}
\mathcal{A}_{\text{SIC}}=\mathcal{A}_{\text{SIC}}'-k\sum_{x}P(b=x\lvert x,\mathbf{povm}),
\end{equation}
for some $k>0$. It follows that the maximal value $\mathcal{A}_{\text{SIC}}=1/2(1+1/\sqrt{3})$ implies that the setting $\mathbf{povm}$ is self-tested as the measurement  $\mathcal{M}_{\text{SIC}}$.

\subsection{Self-testing the trine-POVM }
We consider the second example treated in the main text, namely the so-called trine-POVM.  Here, we provide all details for the discussion in the main text.

In the step 1, we construct a prepare-and-measure experiment in which Alice has a random input $x\in\{1,2,3\}$ and Bob has random input $y\in\{1,2\}$. The aim is to construct a witness $\mathcal{A}_{\text{tri}}'$, the maximal value of which implies Alice's preparation Bloch vectors to form an equilateral triangle on the Bloch sphere. We choose
\begin{equation}
\mathcal{A}'_{\text{tri}}=\sum_{x,y,b}T_{x,y}(-1)^{b}P(b\lvert x,y),
\end{equation}
where $T_{x,1}=\left[1,1,-1\right]$ and $T_{x,2}=\left[\sqrt{3},-\sqrt{3},0\right]$. In a quantum model, the witness can conveniently be written as $
\mathcal{A}'_{\text{tri}}=-1+2\sum_{y} \Tr\left[M_y^0W_y\right]$, where $W_y= \sum_x T_{x,y}\rho_x$. Due to projectors being extremal for binary-outcome measurements, and rank-two projectors leading to trivial statistics for qubits, the optimal measurements are rank-one projectors. Hence, the optimal choice of $M_y^0$ is to align it with the eigenvector of $W_y$ corresponding to its largest eigenvalue. This leads to
\begin{equation}
\mathcal{A}'_{\text{tri}}= -1+2\sum_y \lambda_{\text{max}}\left[W_y\right].
\end{equation}
Since $\rho_1$ and $\rho_2$ can always be viewed to be on a disk of the Bloch sphere, and since $W_1=\rho_1+\rho_2-\rho_3$ and $W_2=\sqrt{3}\rho_1-\sqrt{3}\rho_2$, it is evident that a maximisation of $\mathcal{A}'_{\text{tri}}$ requires also $\rho_3$ to be in that same disk of the Bloch sphere. Without loss of generality, we can choose that disk to be the $xz$-plane. Writing $\rho_x=1/2\left(\openone+\vec{m}_x\cdot \vec{\sigma}\right)$, we may choose $\vec{m}_1=\left[1,0,0\right]$ and parametrise the remaining two preparations of Alice by $\vec{m}_x=\left[\cos \theta_x,0,\sin \theta_x\right]$ for $x\in\{2,3\}$. This leads to
\begin{align}\nonumber
	& \lambda_{\text{max}}\left[W_1\right]= \frac{1}{2}+\frac{1}{2}\sqrt{3+2 \cos (\theta_2)-2 \cos (\theta_2-\theta_3)-2 \cos (\theta_3)}\\\label{qq1}
	& \lambda_{\text{max}}\left[W_2\right]=\sqrt{3}\left\lvert \sin\left(\frac{\theta_2}{2}\right)\right\rvert.
\end{align}
We must now find the values of $(\theta_2,\theta_3)$ that lead to a maximal value of $\mathcal{A}'_{\text{tri}}$. We find the optimal solution,
\begin{equation}\label{qq2}
\frac{\partial \mathcal{A}'_{\text{tri}}}{\partial \theta_3}=0\Rightarrow \theta_3=\pm \arccos\left(\pm \left\lvert \cos\left(\frac{\theta_1}{2}\right)\right\rvert\right).
\end{equation}
However, the two solutions corresponding to a positive sign inside $\arccos$ do not to lead to a maximal value of $\mathcal{A}'_{\text{tri}}$. Therefore, we consider the two solutions that correspond to a negative sign inside $\arccos$. Solving $d\mathcal{A}'_{\text{tri}}/d\theta_2=0$, one obtains the unique solution $\theta_2=\mp 2\pi/3$. This implies $\theta_3=\pm 2\pi/3$, and that the maximal witness value is $\mathcal{A}'_{\text{tri}}=5$. Furthermore, we have that $m_j\cdot m_k=-1/2$ for $j\neq k$ which characterises an equilateral triangle. Thus, a maximal witness value implies that Alice's three preparations form an equilateral triangle in some disk of the Bloch sphere.

In step 2, we supply Bob with an additional measurement $\mathbf{povm}$ which has three outcomes, and thus ensure that a maximal value of
\begin{equation}
\mathcal{A}_{\text{tri}}=\mathcal{A}'_{\text{tri}}-k\sum_{x}P(b=x\lvert x,\mathbf{povm}),
\end{equation}
namely $\mathcal{A}_{\text{tri}}=5$, self-tests the trine-POVM up to unitaries.

\subsection{Self-testing the trine-POVM in a symmetric scenario}\label{secsym}
There is typically no unique prepare-and-measure scenario for completing step 1 (in the main text self-testing method) for any given target measurement. This allows for some freedom in constructing the scenario that is easy to analyse.  We illustrate this possibility here by again considering the trine-POVM in a scenario that exhibits some symmetries. As it turns out, this formulation is particularly handy for certification of non-projective measurements (studied in the next Appendix).

Let Alice have three inputs $x\in\{0,1,2\}$ and Bob have three inputs $y\in\{0,1,2\}$ with binary outcomes $b\in\{0,1\}$. Consider the witness
\begin{equation}\label{tristat}
\mathcal{A}'_{\text{sym}}=\frac{1}{9}\sum_{x,y}P(b=\delta_{x,y}|x,y).
\end{equation}
In a quantum model, we can express this as
\begin{equation}
\mathcal{A}'_{\text{sym}}=\frac{1}{3}+\frac{1}{9}\sum_{y}\Tr\left[M_y^0W_y\right],
\end{equation}
where $W_y=\sum_{x}(-1)^{\delta_{x,y}}\rho_x$. Since we are considering binary measurements on qubits, the optimal measurements are always projective and rank-one. Using the Cauchy-Schwarz inequality followed by applying the concavity inequality \eqref{conc}, we find
\begin{equation}\label{s2}
\mathcal{A}'_{\text{sym}}\leq \frac{1}{3}+\frac{1}{9}\sum_{y} \sqrt{\Tr\left[M_y^0W_y^2\right]}\leq \frac{1}{3}+\frac{1}{3\sqrt{3}} \sqrt{\sum_{y}\Tr\left[M_y^0W_y^2\right]}.
\end{equation}
Evaluating the expression under the square root with the Bloch sphere parameterisation $\rho_x=1/2\left(\openone+\vec{m}_x\cdot\vec{\sigma}\right)$ while optimally aligning $M_y^0$ with $W_y^2$, we obtain
\begin{multline}
\sum_{y}\Tr\left[M_y^0W_y^2\right]\leq \frac{1}{4}\left[3+3\left(|\vec{m}_0|^2+|\vec{m}_1|^2+|\vec{m}_2|^2\right)-2\left(\vec{m}_0\cdot \vec{m}_1+\vec{m}_0\cdot \vec{m}_2+\vec{m}_1\cdot \vec{m}_2\right)+2\sum_{y}\norm{\sum_{x}(-1)^{\delta_{x,y}}\vec{m}_x}\right].
\end{multline}
In order to proceed, we label $\alpha=\left(|\vec{m}_0|^2+|\vec{m}_1|^2+|\vec{m}_2|^2\right)$ and $\beta =-2\left(\vec{m}_0\cdot \vec{m}_1+\vec{m}_0\cdot \vec{m}_2+\vec{m}_1\cdot \vec{m}_2\right)$. Furthermore, we apply the concavity inequality \eqref{conc} to the sum of norms appearing above, and obtain
\begin{equation}\label{s1}
\sum_{y}\Tr\left[M_y^0W_y^2\right]\leq \frac{1}{4}\left[3+3\alpha+\beta+2\sqrt{3}\sqrt{3\alpha+\beta}\right]=\frac{1}{4}\left(\sqrt{3\alpha+\beta}+\sqrt{3}\right)^2.
\end{equation}
Hence, we need only to maximise the expression $3\alpha+\beta$. It is immediately clear that both $\alpha$ and $\beta$ are maximal for pure states, i.e., choosing
$|\vec{m}_x|=1$. This gives $\alpha=3$. In order to maximise $\beta$, we must choose the three preparation Bloch vectors to be co-planar. Due to the freedom of applying a global unitary, we can choose $\vec{m}_0=\left[1,0,0\right]$, $\vec{m}_1=\left[\cos\theta,0,\sin\theta\right]$. and  $\vec{m}_2=\left[\cos\phi,0,\sin\phi\right]$. Consequently, we obtain
\begin{equation}
\beta=2\left(\cos \theta+\cos\left(\theta-\phi\right)+\cos\phi\right).
\end{equation}
The equation $\partial\beta/\partial\theta=0$ has four solutions of which two correspond to the maximum of $\beta$. These are $\theta=\pm \arccos\left(-\left|\cos\left(\frac{\phi}{2}\right)\right|\right)$. The corresponding optimum $\beta=3$ is found at $\theta=\mp 2\pi/3$. Re-inserting this into Eq.\eqref{s1}, the right-hand-side becomes $27/4$, which re-inserted into Eq.\eqref{s2} returns the upper bound
\begin{equation}
\mathcal{A}'_{\text{sym}}\leq 5/6.
\end{equation}
This upper bound is tight, as it can be saturated with an explicit quantum strategy. Furthermore, it follows that a maximal quantum value of $\mathcal{A}'_{\text{sym}}$ implies that Alice's preparations correspond to three pure co-planar and equiangular Bloch vectors, i.e., they form an equilateral triangle since $m_j\cdot m_k=-1/2$ for $j\neq k$.

Hence, by step 2, we supply Bob with another measurement setting $\mathbf{povm}$ with three outcomes, and therefore know that a maximal value of
\begin{equation}
\mathcal{A}_{\text{tri2}}=\frac{1}{9}\sum_{x,y=0,1,2}P(b=\delta_{x,y}|x,y)-k\sum_{x} P(b=x|x,\mathbf{povm}),
\end{equation}
self-tests the trine-POVM up to unitaries whenever $k>0$.

\section{Certification of non-projective measurements based on the trine-POVM}\label{ExamplesCertification}

In the main text, we evaluated a bound for projective measurements in a prepare-and-measure scenario based on the qubit SIC-POVM. Here, we do the same for the prepare-and-measure scenario of section~\eqref{secsym} based on the trine-POVM.

Due to the many apparent symmetries of $\mathcal{A}_{\text{sym}}$, the evaluation of a projective bound is greatly simplified. We can without loss of generality assign the two non-trivial projectors in the measurement $\mathbf{povm}$ to the outcomes $b\in\{0,1\}$, and the zero-projector to outcome $b=2$. Then, we construct the observable $M_3\equiv M_{\mathbf{povm}}=M_{\mathbf{povm}}^0-M_{\mathbf{povm}}^1$. An evaluation of the witness in terms of observables gives
\begin{align}
\mathcal{A}_{\text{sym}}&=\frac{1-2k}{2}+\frac{1}{18}\sum_{x,y}(-1)^{\delta_{x,y}}\Tr\left(\rho_xM_y\right)-\frac{k}{2}\sum_{x=0,1}(-1)^x\Tr\left(\rho_xM_3\right)\\
& =\frac{1-2k}{2}+\frac{1}{18}\Big(\Tr\left[\rho_0\left(-M_0+M_1+M_2-9kM_3\right)\right]+\Tr\left[\rho_1\left(M_0-M_1+M_2+9kM_3\right)\right]+\Tr\left[\rho_2\left(M_0+M_1-M_2\right)\right]\Big).
\end{align}
We apply the Cauchy-Schwarz inequality while using that the optimal projective measurements are rank-one and therefore can be associated to observables of the form $M_y=\vec{n}_y\cdot\vec{\sigma}$. We obtain
\begin{equation}
\mathcal{A}_{\text{sym}}\leq\frac{1-2k}{2}+\frac{1}{18}\Big(\sqrt{r_k+2\left(-1,-1,9k,1,-9k,-9k\right)\cdot \vec{N}}+\sqrt{r_k+2\left(-1,1,9k,-1,-9k,9k\right)\cdot \vec{N}}+\sqrt{3+2\left(1,-1,-1\right)\cdot\vec{M}}\Big),
\end{equation}
where $r_k=3+81k^2$, and $\vec{N}=\left(\vec{n}_0\cdot \vec{n}_1,\vec{n}_0\cdot \vec{n}_2,\vec{n}_0\cdot \vec{n}_3,\vec{n}_1\cdot \vec{n}_2,\vec{n}_1\cdot \vec{n}_3,\vec{n}_2\cdot \vec{n}_3\right)$, and $\vec{M}=\left(\vec{n}_0\cdot \vec{n}_1,\vec{n}_0\cdot \vec{n}_2,\vec{n}_1\cdot \vec{n}_2\right)$. Subsequently, we apply the concavity inequality \eqref{conc} to the first two square-root expressions and obtain
\begin{equation}
\mathcal{A}_{\text{sym}}\leq \frac{1-2k}{2}+\frac{\sqrt{2}}{18}\sqrt{2r_k-4\vec{n}_0\cdot\vec{n}_1+36k\vec{n}_3\cdot \left(\vec{n}_0-\vec{n}_1\right)}+\frac{1}{18}\sqrt{3+2\vec{n}_0\cdot \vec{n}_1-2\vec{n}_2\cdot\left(\vec{n}_0+\vec{n}_1\right)}.
\end{equation}
The optimal choice of $\vec{n}_3$ is to align it with the vector $\vec{n}_0-\vec{n}_1$. Similarly, the optimal choice of $\vec{n}_2$ is to anti-align it with the vector $\vec{n}_0+\vec{n}_1$. Labelling $x=\vec{n}_0\cdot\vec{n}_1$, we find
\begin{eqnarray}
\mathcal{A}_{\text{sym}}\leq \frac{1-2k}{2}+\frac{\sqrt{2}}{18}\sqrt{2r_k-4x+36k\sqrt{2}\sqrt{1-x}}+\frac{1}{18}\sqrt{3+2x+2\sqrt{2}\sqrt{1+x}}\equiv f_k(x).
\end{eqnarray}
Notice that we have reduced the original problem to depend only on a single real parameter. The optimal choice, $x^*$, is found from the standard method of solving $d f_k(x)/d x=0$ by suitable means. The bound on projective measurements is then given by $\mathcal{B}(k)\equiv f_k(x^*)$, i.e.,
\begin{equation}
\mathcal{A}_{\text{sym}}\stackrel{\text{Projective}}{\leq} \mathcal{B}(k).
\end{equation}

To consider the tightness of this bound, we have numerically optimised $\mathcal{A}_{\text{sym}}$ under the constraint of unit-trace measurement operators, which includes projective rank-one operators as a special case. We performed the optimisation for $k=1/100,2/100,\ldots,1$ and (up to machine precision) saturated the bound $\mathcal{B}(k)$ in all cases.

\section{Experimental details} \label{ExperimentDetails}

\subsection{Errors}
{\em Statistical uncertainties.--}%
Because of the relatively high rates in our experiments (approximately $10^4$ events per second), around one to two hours of measuring were enough to have statistical uncertainties smaller than systematic ones. Assuming a Poissonian distribution on the number of counts $N$, the statistical uncertainty was taken to be $\sqrt{N}$, and propagated on the calculated results.

{\em Systematic errors.--}%
Here we evaluate the most important sources of systematic errors in our setup:

(1) Motor precision. Because the certificate of non-projective measurements only relies on the final estimation of the witness in the prepare-and-measure scenario, and not on the particular states or measurement settings employed, how accurate the motors are in setting each wave plate is effectively irrelevant. However,  whenever one wave plate setting is repeated during an experiment (i.e., corresponding to one and the same setting allowed in the prepare-and-measure scenario), the precision with which the motors are able to do this directly propagates into un uncertainty on the repeated setting, and therefore on the measured probabilities used to estimate $\mathcal{A}_{\text{SIC}}$ or $\mathcal{A}_{\text{tri}}$.

The motors used in the experiments have a precision equal to $0.02 \degree$. In order to derive an uncertainty, Monte Carlo simulations with $10^5$ runs, with wave plate setting distributed normally around set values with FWHM of $0.02 \degree$, were performed.
The resulted uncertainties are slightly greater than the statistical counterparts, and constitute the sole contribution to the systematic errors reported in the experimentally obtained witness value in the prepare-and-measure game.

(2) Detector dark counts. The single-photon avalanche photo-diodes employed in the protocols have dark count rates of approximately 500 counts per second. A coincidence between a true trigger detection and a dark count constitutes a ``wrong'' count in our results. However, with the numbers mentioned above, such rates can be calculated to be of the order of $10^{-10}$ events per second, and are therefore negligible.

\end{document}